\documentclass[
aip,
amsmath,amssymb,
reprint,
]{revtex4-1}

\usepackage{graphicx}
\usepackage{dcolumn}
\usepackage{bm}
\usepackage{color,soul} 
\usepackage{textcomp}
\usepackage[utf8]{inputenc}
\usepackage[T1]{fontenc}
\usepackage{mathptmx}
\usepackage{etoolbox}
\usepackage{tabularx}
\usepackage{stackengine}
\usepackage{empheq}
\usepackage{xcolor}
\usepackage{subcaption}
\usepackage{placeins}
\usepackage{hyperref}

\newcommand*{\colorboxed}{}
\def\colorboxed#1#{%
  \colorboxedAux{#1}%
}
\newcommand*{\colorboxedAux}[2]{%
  \begingroup
    \colorlet{cb@saved}{.}%
    \color#1{cyan}%
    \boxed{%
      \color{cb@saved}%
      #2%
    }%
  \endgroup
}

\makeatletter
\def\@email#1#2{%
	\endgroup
	\patchcmd{\titleblock@produce}
	{\frontmatter@RRAPformat}
	{\frontmatter@RRAPformat{\produce@RRAP{*#1\href{mailto:#2}{#2}}}\frontmatter@RRAPformat}
	{}{}
}%
\makeatother
\begin{document}

	\title[]{Density and Particle Sourcing Optimization in a Helicon Plasma Source Prototype For Wakefield Accelerator Applications}
	
	\author{M. Zepp}
	\email{mzepp@wisc.edu}
	\author{M. Granetzny}
	\author{O. Schmitz}
	\affiliation{Department of Nuclear Engineering and Engineering Physics, University of Wisconsin--Madison, Madison, Wisconsin 53706, USA}
	
	\date{\today}
	
	\begin{abstract}
        Helicon plasmas are being considered as plasma sources for wakefield accelerators, subject to strict density requirements. We present various mechanisms to increase axial density homogeneity in a helicon plasma for implementation in such an accelerator. We consider various background neutral flow configurations for helicons generated with first one antenna and then two identical antennas in a $2$ meter long, $52$ mm diameter plasma chamber with homogeneous magnetic field. In the case of a single antenna, the ionization source rate and density profiles are not significantly influenced by the background neutral flow. The use of a second antenna expands the plasma axially along the device, and results in an increased dependence of the axial density profile on the background neutral flow. We find an increase in axial homogeneity by a factor of two when there is no background neutral flow compared to when there is flow in either direction relative to the plasma. The minimum axial density deviation accomplished by optimization of RF antenna and neutral flow was $5\%$. This is still a factor of $20$ above the nominal homogeneity requirement, but means to further improve this have been established in this study.
	\end{abstract}
	
	\maketitle
    
    \section{Introduction}\label{sec:intro}
	Helicon plasmas are high density, low temperature plasmas generated by bounded whistler waves. These plasmas have potential applications in space thrusters,\cite{Saini_2024} current drive in magnetic confinement fusion energy systems,\cite{Tooker_2017,Pinsker_2024} and plasma sources for next-generation particle accelerators.\cite{Stollberg_2024} The Advanced Proton Driven Plasma Wakefield Acceleration Experiment (AWAKE) at CERN has proposed using helicon plasmas in wakefield accelerators with length scales of $100$ m or longer.\cite{Muggli_2020} Such a helicon would be produced by a series of RF antennas located along the plasma length. The optimized configuration is currently unclear. The electric wake field in the plasma for the AWAKE linear accelerator would be driven by proton bunches from the Super Proton Synchrotron (SPS) in the CERN beam complex. These proton bunches are fairly long at $6-12$ cm, and carry $19$ kJ per bunch. The bunch length leads to the requirement that the plasma density along the length of the bunch have axial variations of at most $0.25\%$ at densities of at least $10^{21}$  m$^{-3}$.\cite{Muggli_2018} To achieve this density uniformity requirement, the device setup and input parameters must be optimized. In particular the antenna shape, location, and power as well as the neutral particle fueling terms, like neutral pressure and gas valve locations are of interest.\\
    
    The Madison AWAKE Prototype (MAP)\cite{Granetzny_2025} at the University of Wisconsin--Madison was designed to enable these optimization studies. In previous studies, we have shown how the magnetic field direction and antenna helicity influence the plasma launch direction,\cite{Granetzny_2023} and we have presented a method for measuring the $2$D axisymmetric ionization source rate profile, concluding that this ionization source rate is dominated by radial fueling in MAP.\cite{Zepp_2024}  In this article, we present studies to identify key controls to reach an axial density profile with the necessary uniformity. We consider various neutral fueling schemes and compare the fueling in one-antenna and two-antenna helicons in a two meter long chamber with a homogeneous magnetic field. In Section \ref{sec:apparatus}, we describe the MAP helicon device and relevant diagnostics for this study. In Section \ref{sec:fueling} we describe the fueling configurations that we utilize in this work. In Section \ref{sec:oneant}, we present $2$D density, ionization source rate and axial momentum balance results, and we explore how neutral fueling influences the plasma profiles. In Section \ref{sec:twoant}, we continue this analysis with a second antenna and present results regarding the coupling between antennas. We conclude by identifying the configurations with no background neutral flow as strongly preferable for wakefield accelerator applications.\\

	    \section{Experimental Apparatus}\label{sec:apparatus}
\begin{figure}[htbp]
		\centering
		\includegraphics[width=\linewidth]{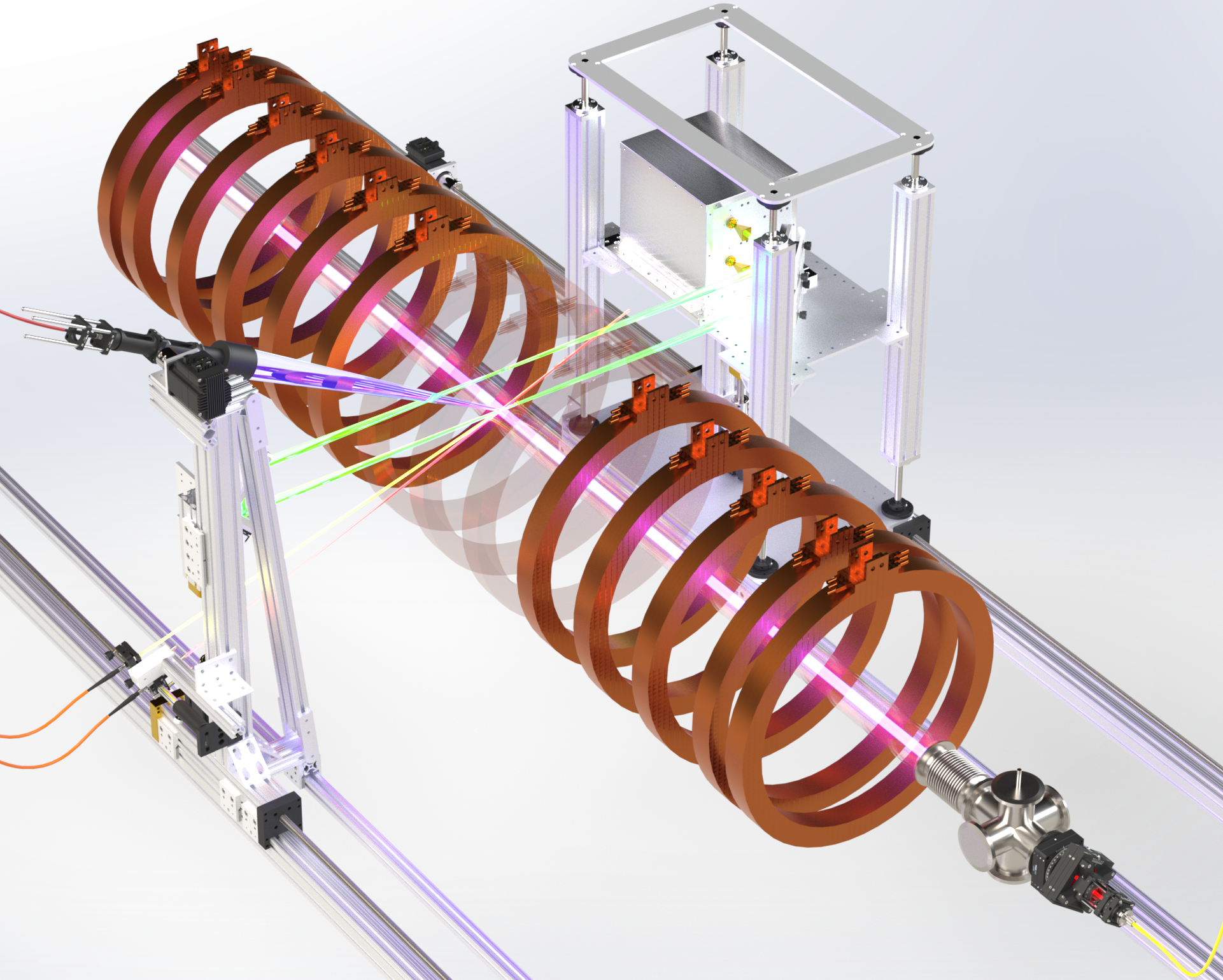}
		\caption{Key features of the Madison AWAKE Prototype (MAP) experiment, showing the lines of sight of the interferometer (green), passive spectrometer (yellow), LIF laser (red), and LIF collection optics (blue). The central magnetic field coils are depicted as transparent for easier viewing of the central plasma region.}
		\label{fig:MAPCAD}
	\end{figure}

    The research presented in this work was conducted on the Madison AWAKE Prototype (MAP).\cite{Granetzny_2025}  A CAD rendering of key MAP components, including diagnostics, is shown in Figure \ref{fig:MAPCAD}. The main plasma chamber is a $2.15$ m long borosilicate glass tube with inner and outer diameters of $52$ and $56$ mm, respectively. Two identical $10$ cm long antennas can be used to excite helicon plasmas propagating along the 50 mT magnetic field produced by a set of $14$ electromagnets as seen in Figure \ref{fig:MAPCAD}. The antennas were optimized for efficient RF wave coupling \cite{Granetzny_2023,Granetzny_2025} and they are designed to accommodate a $1$ cm air gap between the glass surface and the antenna for improved cooling. This gap prevents excessive thermal gradients that have previously led to cracks in the glass. Each antenna is powered by its own RF generator capable of delivering up to $10$ kW of RF power at $13.56$ MHz, either pulsed or continuous. This yields reliable generation of helicon plasmas in the high $10^{19}$ m$^{-3}$ range at $\approx3-5$ kW and in the $10^{20}$ m$^{-3}$ for powers of $>6$ kW.  MAP and its capabilities are described in detail in\cite{Granetzny_2025}.\\

    A microwave interferometer on MAP has a cutoff density of $1.4\times10^{20}$ m$^{-3}$ and a practical density measurement limit of $8\times10^{19}$ m$^{-3}$ due to refraction effects at the plasma for this microwave frequency. It provides absolute, line-averaged density measurements on MAP and is described in detail in \cite{MAPInterferometer}. The interferometer housing is located on the back side of MAP, in the upper right of Figure \ref{fig:MAPCAD}, and the return mirrors are positioned directly opposite the housing. Two spectrometers provide passive spectral measurements of MAP plasmas. The collection fiber and lens for the spectrometers are located at the lower left of Figure \ref{fig:MAPCAD}. The main diagnostic used on MAP is the laser-induced fluorescence (LIF) system, which provides non-perturbative, active measurements of ion temperatures, velocities, and relative densities. All diagnostics can measure the same position of the plasma simultaneously, though only one spectrometer can be used at a time.\\

    The LIF system on MAP is the same system that was previously used on the MARIA helicon device,\cite{Green_2019,Green_2020,Green_2020B} and follows the design of Severn et al.\cite{Severn_1998} The system was transferred to MAP in $2022$, where it has been used for density and particle balance measurements.\cite{Granetzny_2023,Zepp_2024} We use laser tuning to pump the $3$d$^4$F$_{7/2}$ to $4$p$^4$D$_{5/2}$ argon ion transition at $668.614$ nm. Fluorescence light at $442.6$ nm is collected and analyzed to extract density and flow measurements.\\

    The LIF laser can be injected radially or axially to measure flow velocities in either direction. The radial and axial injection optics are both visible in Figure \ref{fig:MAPCAD}, with the axial injection located at the bottom right of the figure, and the radial injection near the bottom left. The collection optics are located at the mid-upper left of the figure. For radial injection, the laser polarization is in the direction of the magnetic field, so that a $\pi$ transition is excited, in which the z-projection of the angular momentum does not change. For axial injection, the $\pi$ transitions are not accessible, so $\sigma^\pm$ transitions, defined by a change of the quantum number $m$, are excited by using circularly polarized light. The methodology presented by Green et al.\cite{Green_2019} was used to extract axial ion velocities from the average of the velocities extracted from the $\sigma^+$ and $\sigma^-$ transitions.\\

    For measurements of the density with LIF, we use the scaling law derived by Green\cite{Green_2020} and Zepp\cite{Zepp_Thesis} on the MARIA device, and corrected by a scale factor derived with the interferometer on MAP. This scaling law is seen to apply consistently at RF powers below $3$ kW, but behavior at higher powers is inconsistent, likely due to depletion of the pumped state. Density measurements at higher power require an upgrade to the MAP interferometer or a separate diagnostic. The LIF intensity $S$, calculated as the product of the width and amplitude of the measured velocity distribution function, is related to the electron density $n$ in m$^{-3}$ as 
    \begin{equation}
        n = \left(1.85\times 10^{18}\right)\left(\frac{S}{1\ \mu \text{V s}^{-1}}\right)^{0.3}.
    \end{equation}
    This scaling law was observed to apply at electron temperatures below $6$ eV, with density measurements from $2\times10^{17}-2\times10^{19}$ m$^{-3}$ made with an RF compensated Langmuir probe and an interferometer.\cite{Sun_2004,Green_2020,Zepp_Thesis}\\
    
    \section{Fueling Schemes in a Helicon Plasma}\label{sec:fueling}

\begin{figure}[htbp]
    	\centering
    	\begin{subfigure}{\columnwidth}
    		\centering
    		\includegraphics[width=\linewidth]{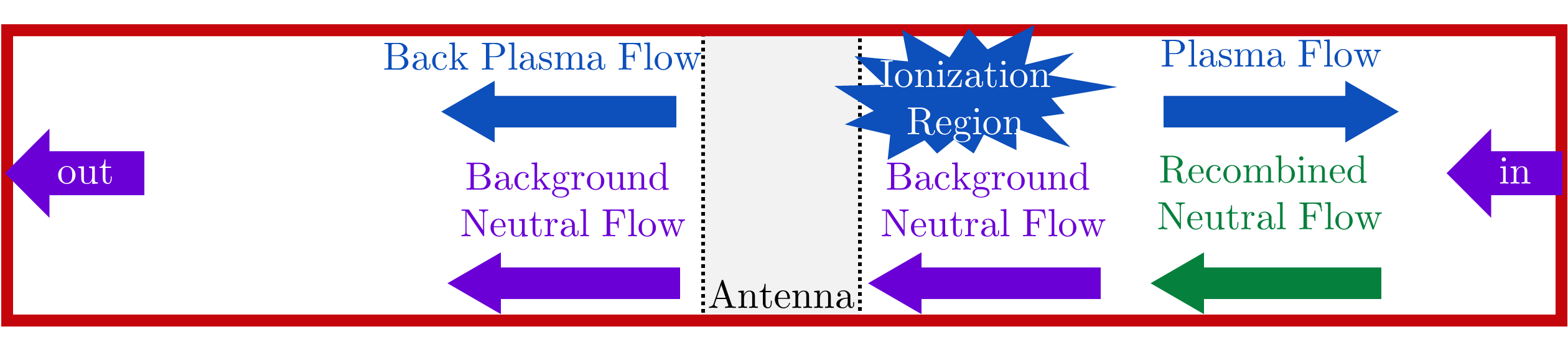}
    		\caption{Antiparallel flow configuration, $\Gamma_n\leftarrow\ |\ \Gamma_p\rightarrow$}
    		\label{fig:FlowA}
    	\end{subfigure}
    	\begin{subfigure}{\columnwidth}
    		\centering
    		\includegraphics[width=\linewidth]{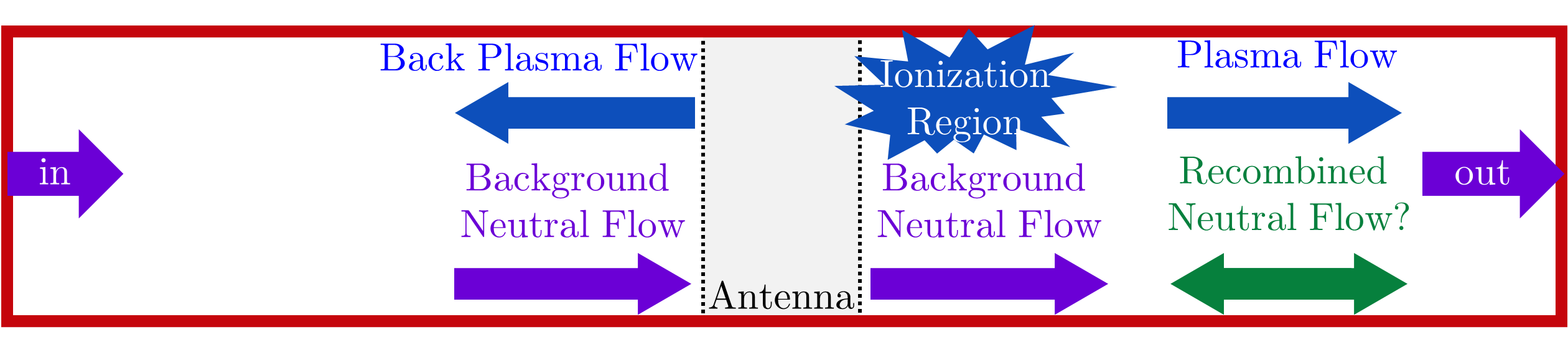}
    		\caption{Parallel flow configuration, $\Gamma_n\rightarrow\ |\ \Gamma_p\rightarrow$}
    		\label{fig:FlowB}
    	\end{subfigure}
        \begin{subfigure}{\columnwidth}
            \centering
            \includegraphics[width=\linewidth]{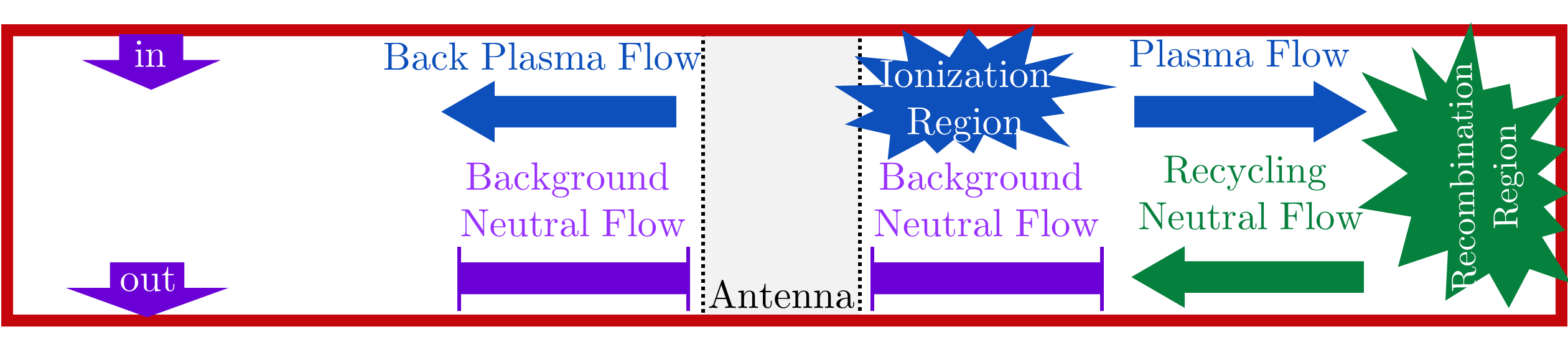}
            \caption{No-flow configuration, $\Gamma_n = 0 \ |\ \Gamma_p\rightarrow$}
            \label{fig:FlowC}
        \end{subfigure}
    	\caption{Possible flow schematics for three different neutral and plasma flow configurations. Neutral particles are injected and pumped out (purple) in three different configurations. This background flow exists in the absence of a plasma. In all three configurations, the principal plasma flow (blue) is from the antenna to the right. Any flow of the recombined neutrals necessary to maintain the axial flux balance is indicated (green)}\label{fig:Flows}
    \end{figure}

    One characteristic of helicons that we can exploit to optimize the fueling scheme is the directionality of the plasma, which we explained in a previous publication.\cite{Granetzny_2023} A density peak occurs to one side of the antenna in a MAP-like single-antenna helicon experiment, i.e. a setup in which the magnetic field to both sides of the antenna is comparable and there is sufficient space on either side for the wave to propagate. We define the direction from the antenna towards the density peak as the downstream plasma direction, denoted $\Gamma_p$. This is represented in Figure \ref{fig:Flows} by the blue arrow pointing to the right, labeled "Plasma flow." With the plasma flow pointing to the right ($\Gamma_p\rightarrow$), we consider three possibilities for background neutral flow, which we define as the direction in which neutrals flow in the absence of plasma. This direction is determined by the gas injection and pumping scheme of the device. Defining the background neutral flow direction as $\Gamma_n$, we may have the plasma and neutral flows antiparallel ($\Gamma_n\leftarrow|\Gamma_p\rightarrow$) or parallel ($\Gamma_n\rightarrow|\Gamma_p\rightarrow$), or there may be no neutral flow ($\Gamma_n=0|\Gamma_p\rightarrow$). These three cases are indicated in Figures \ref{fig:FlowA}, \ref{fig:FlowB}, and \ref{fig:FlowC}, respectively. We refer to these configurations as the antiparallel flow configuration, parallel flow configuration, and no-flow configuration, respectively.\\

    \section{One Antenna Helicon Operation}\label{sec:oneant}
    With a single antenna in MAP at $1$ kW, $47$ mT, and $3$ Pa, we calculate the $2$D density and ionization source rate profiles from the ion continuity equation based on  LIF measurements following the techniques described in our previous letter.\cite{Zepp_2024} The ionization source rate $S$ is calculated from the divergence of the ion flux. We take the plasma to be axisymmetric, so the ionization source rate values presented here are calculated by
    \begin{equation}\label{equation:partbal}
        S = \frac{\partial (nV_z)}{\partial z}+\frac{1}{r}\frac{\partial (rnV_r)}{\partial r},
    \end{equation}
    where density $n$, radial velocity $V_r$, and axial velocity $V_z$ are all measured by LIF. We refer to the individual axial and radial contributions as $S_Z$ and $S_R$, respectively, such that 
    \begin{align}
        S_Z &= \frac{\partial (nV_z)}{\partial z}\\
        S_R &= \frac{1}{r}\frac{\partial (rnV_r)}{\partial r}\\
        S &= S_Z + S_R
    \end{align}
    Further details of the method are published in previous publications.\cite{Zepp_2024}\cite{Green_2020}\cite{Zepp_Thesis}\\

    Three configurations of the MAP gas system were used for the comparison of ionization source rate profiles. The antiparallel flow configuration was established by launching the plasma to the right in Figure \ref{fig:MAPCAD} and flowing neutrals from right to left, as depicted in Figure \ref{fig:FlowA}. When setting MAP for the parallel flow configuration, both the neutrals and plasma flow to the left in Figure \ref{fig:MAPCAD}. However, for ease of comparison we show all data axially mirrored around the antenna location, such that the setup can be represented as shown in Figure \ref{fig:FlowB}. The plasma launch direction was changed by changing the direction of the magnetic field while maintaining all other parameters. The mechanism by which the flipped magnetic field changes the launch direction of the plasma is described in detail in our previous article.\cite{Granetzny_2023} The no-flow configuration was established by injecting and pumping argon gas on the left side of the device in Figure \ref{fig:MAPCAD}. In this case, the plasma was launched to the right.\\
    
	\begin{figure}[htbp]
		\centering
		\includegraphics[width=\linewidth]{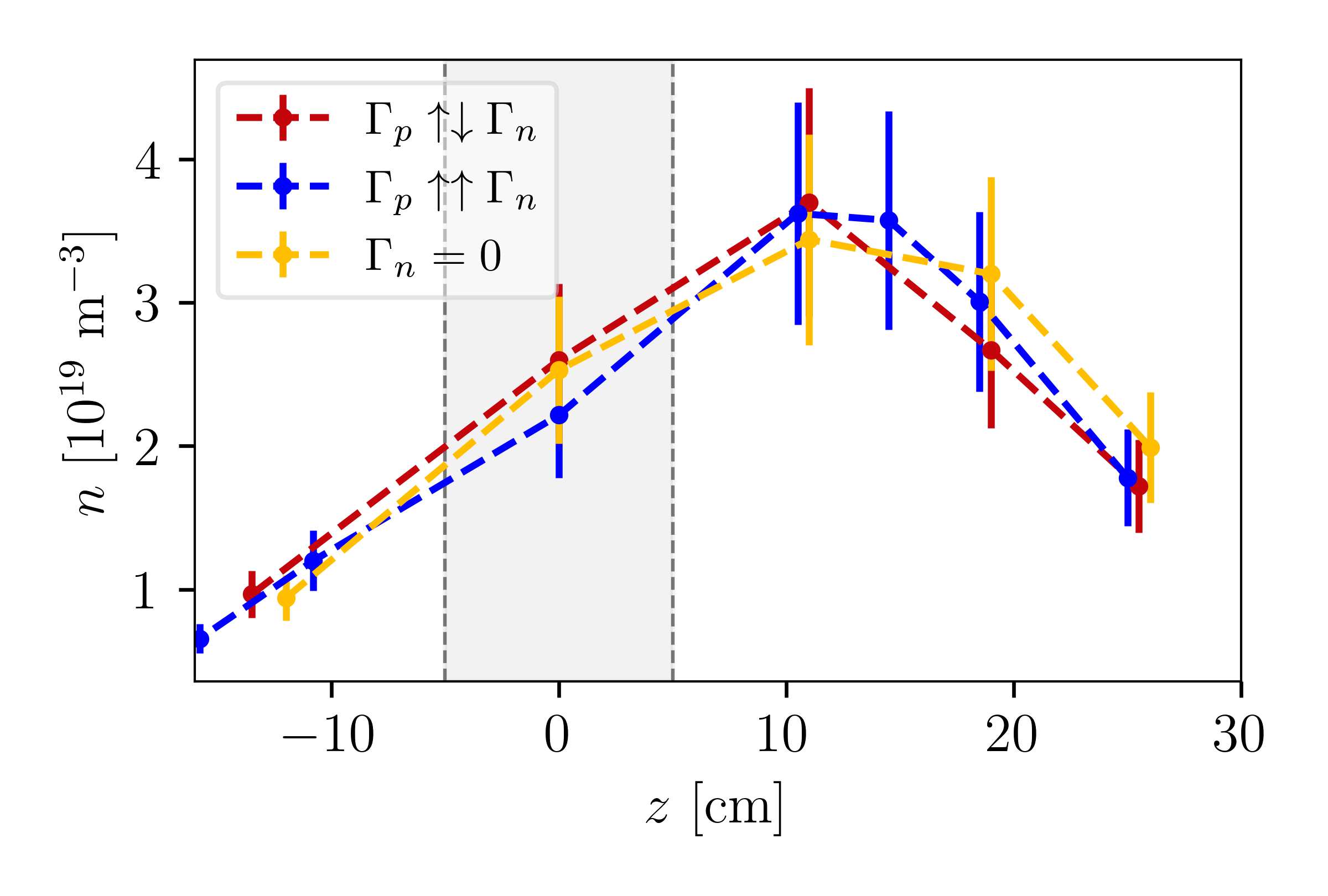}
		\caption{The axial plasma density profile is plotted for the antiparallel flow configuration (red), parallel flow configuration (blue), and no-flow configuration (yellow). The antenna is located in the shaded region.}
		\label{fig:densityAll}
	\end{figure}

	\begin{figure*}[t]
		\centering
		\includegraphics[width=0.9\linewidth]{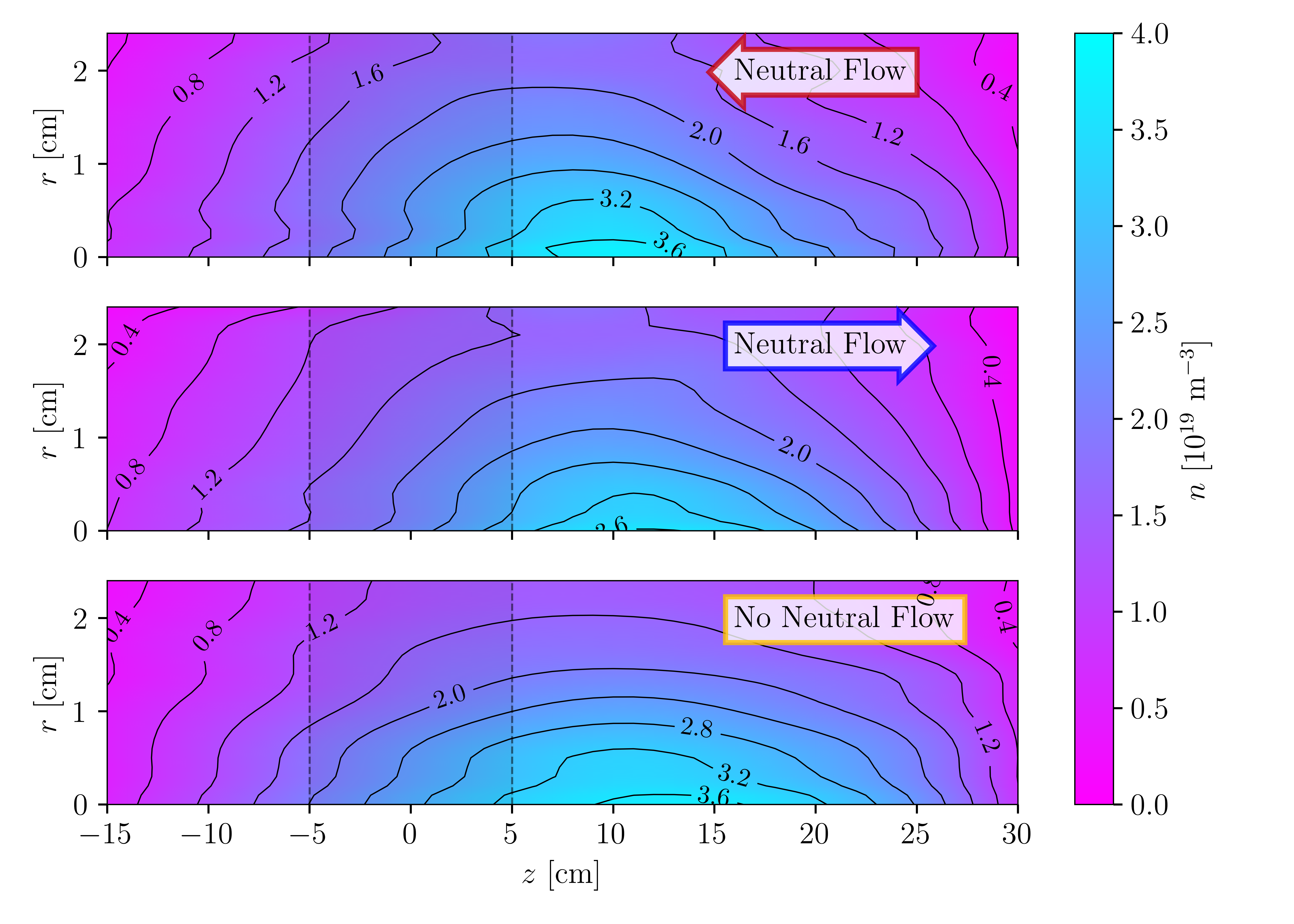}
		\caption{The $2$D density profile is plotted for three plasmas with identical conditions except that the background neutral flow is from right to left (top), background neutral flow is from left to right (middle), and there is no background neutral flow (bottom). }
		\label{fig:densitycontourAll}
	\end{figure*}
    
    The on-axis density profiles are plotted for all three configurations in Figure \ref{fig:densityAll}. The $10$ cm antenna is denoted in the figure by the shaded region. A clear density peak is visible in all three configurations, roughly $12$ cm from the center of the antenna.\\
    
    The differing neutral conditions have very little impact on the density profile. While the only differences are within the errorbars, we note that the antiparallel (red) and parallel (blue) flow configurations are offset slightly axially, and the no-flow configuration (yellow) is slightly more uniform, with a broader but slightly lower density profile. The offset in the antiparallel and parallel flow configurations is small, but, intuitively, the density profile could shift in the direction of the background neutral flow, as is seen in the figure, if the neutrals drag the plasma along. The uncertainty indicated by the errorbars is due primarily to the uncertainty in the LIF density scaling law. The uncertainty in the LIF signal measurements is typically under $1$\%, while the density uncertainty is typically about $20$\%. \\
    
    The full $2$D density profiles for the same three configurations are plotted in Figure \ref{fig:densitycontourAll}. The density peaks are still apparent near $z=12$ cm and the density is notably peaked on-axis. This is in contrast to previous work on MAP, in which we found a slightly hollow profile at $1.3$ kW RF power and $1$ Pa background fill argon pressure.\cite{Granetzny_2025} The higher pressure of $3$ Pa used in this present study leads to adequate argon available to allow particles to penetrate to the axis of the device where they can be ionized and sustain the plasma. \\
        
	\begin{figure*}[t]
		\centering
		\includegraphics[width=0.9\linewidth]{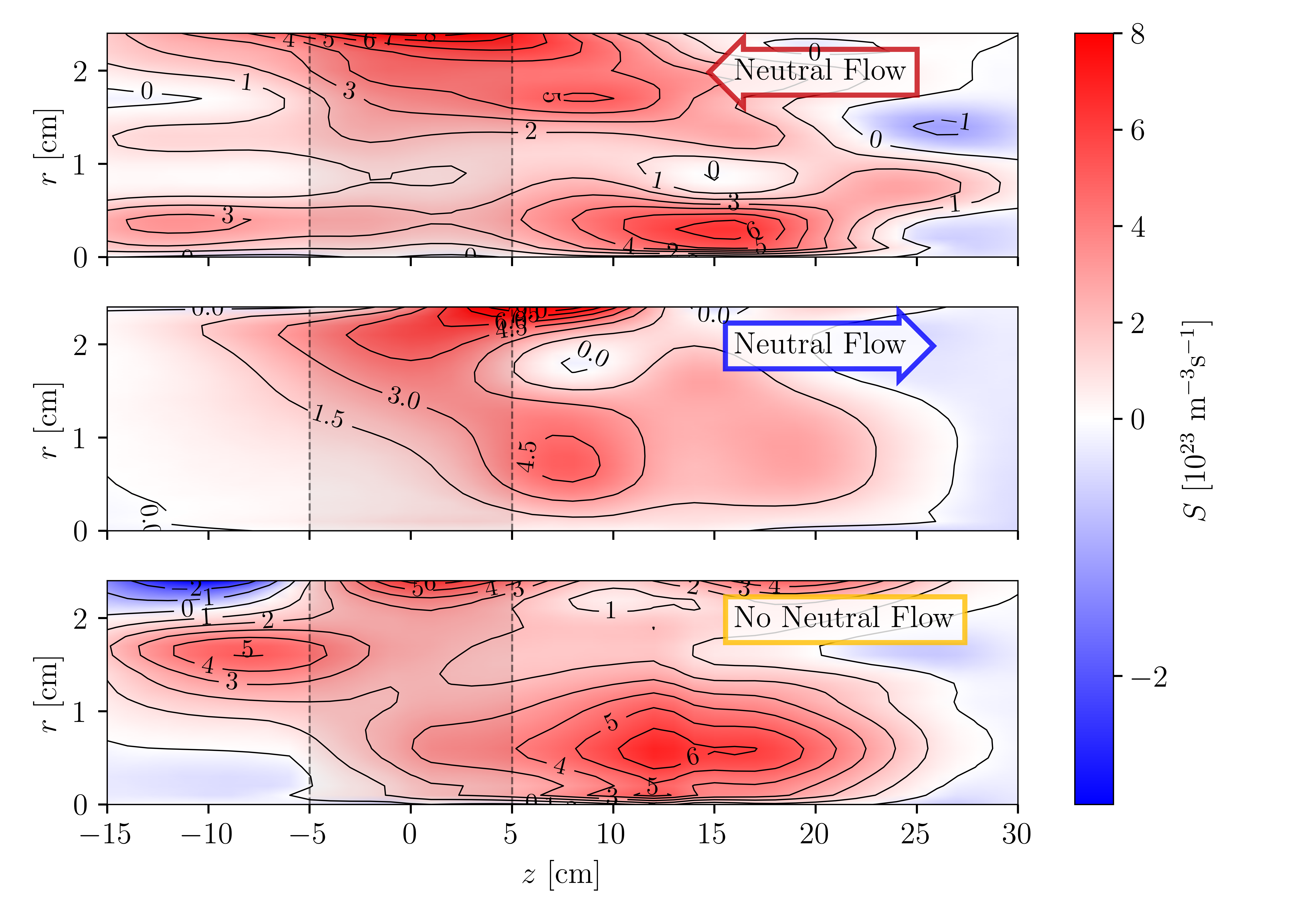}
		\caption{The ionization source rate $S$ is plotted for three plasmas with identical conditions except that the background neutral flow is from right to left (top), background neutral flow is from left to right (middle), and there is no background neutral flow (bottom). These are the same plasmas for which density profiles are plotted in Figure \ref{fig:densitycontourAll}}
		\label{fig:SRAll}
	\end{figure*}

    Our previous findings\cite{Zepp_2024} suggested that the ionization source rate in a MAP-like device is determined primarily by radial fluxes. For this reason, we used a much higher radial resolution than axial resolution in our data acquisition. Axially, we use a resolution of roughly one measurement every $10$ cm, while radially, we use a resolution of roughly one measurement every $2.5$ mm. This radial resolution allows for the measurement of the ionization source rate from Equation \ref{equation:partbal}. \\
    
    The results of the ionization source rate measurements for all three configurations are shown in Figure \ref{fig:SRAll}. The top panel is the ionization rate $S$ for the antiparallel flow configuration, the middle panel is for the parallel flow configuration, and the bottom panel is for the no-flow configuration. Typical uncertainties are roughly $30$\%, although we note that a suspected inhomogeneity in the glass chamber caused larger uncertainty for measurements from $r=0.5-2$ cm roughly $10$ cm upstream from the antenna ($-10$ cm) in the antiparallel flow and no-flow configurations, and the same distance downstream from the antenna ($+10$ cm) in the parallel flow configuration.\\ 

    The ionization source rate profiles are similar for all three configurations. They all exhibit a strong ionization source directly under the antenna near the device boundary and another strong ionization source near the axis in the vicinity of the density peak ($z\approx10$ cm). This is in agreement with previous results for a MAP plasma at a slightly higher power of 1.3 kW and a lower pressure of 1 Pa.\cite{Zepp_2024} As in our past work, we conclude that the strong ionization source rate at the vacuum vessel wall near the antenna is due to the Trivelpiece-Gould mode, and the strong ionization source rate near the axis is due to the helicon mode. The neutral flow does not significantly alter the propagation of the modes, nor does it significantly influence the fueling of the plasma, as suggested by the similarity of the ionization source rate profiles.\\
    
	\begin{figure*}[t]
		\centering
		\includegraphics[width=0.9\linewidth]{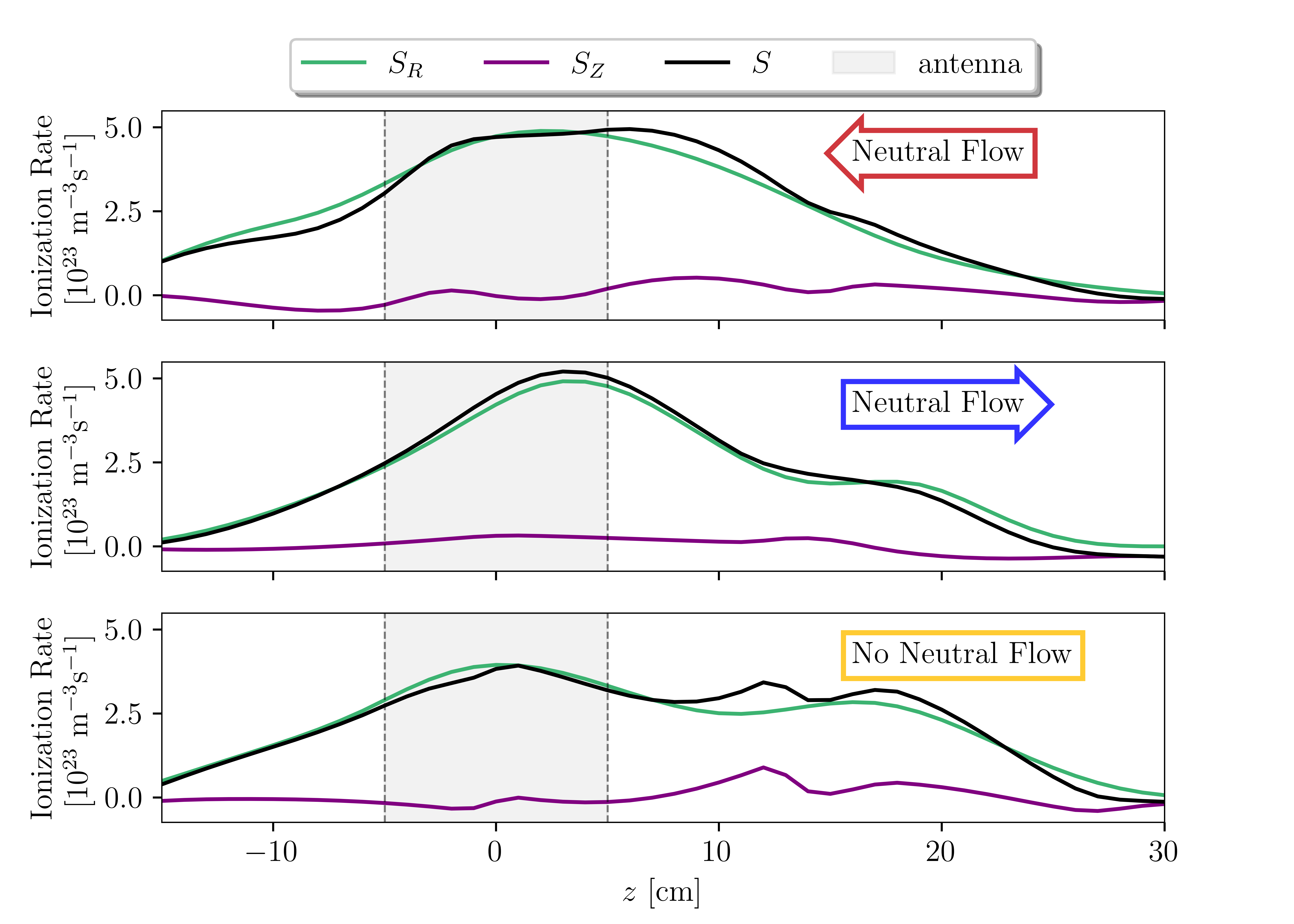}
		\caption{The ionization source rate components $S_R$, $S_Z$, and $S$ are plotted for three plasmas with identical conditions except that the background neutral flow is from right to left (top), background neutral flow is from left to right (middle), and there is no background neutral flow (bottom). Each source rate parameter is averaged over all radii at each axial position.}
		\label{fig:SRcompsAll}
	\end{figure*}

    To understand why the neutral flow does not significantly alter the density or ionization source rate profiles, we examine the radial and axial flux contributions to the total ionization source rate. We found in our previous work that the radial flux contribution makes up roughly $90$\% of the total ionization source rate. To verify that this also applies in the conditions presented here, we plot the individual components $S_R$ and $S_Z$, and the total ionization source rate $S$ for all three configurations in Figure \ref{fig:SRcompsAll}. We see in all three cases that the radial flux contribution $S_R$ vastly exceeds the axial flux contribution $S_Z$ to the total ionization rate $S$, as $S_R$ makes up at least $75\%$ and typically $\approx 90\%$ of $S$ in the peak density region, and $99\%$ of $S$ overall for all three configurations. \\

    We now investigate a momentum balance to consider how momentum transfer could influence the density profile. We use the equation for a 2D axisymmetric axial momentum balance from Stangeby\cite{Stangeby_2000}:
    \begin{equation}\label{equation:MomBalance1}
    	\begin{split}
    		\frac{\partial}{\partial z}\left(p_i + p_e + nm_i v_z^2 + \pi_i\right) & = S_{mom}(r,z)  \\
    		& + \frac{1}{r}\frac{\partial}{\partial r}r\left[m_i v_z \left(D_\perp \frac{\partial n}{\partial r} + n v_{pinch}\right)\right]\\
    		& + \frac{1}{r}\frac{\partial}{\partial r}r \left[\eta_\perp \frac{\partial v_z}{\partial r}\right].
    	\end{split}
    \end{equation}
    Here $p_i$ and $p_e$ are ion and electron pressure, respectively, $\pi_i$ is the viscous stress, $S_{mom}$ is the momentum source rate, which is typically a sink due to friction with neutrals, $D_\perp$ is the cross-field diffusion coefficient, $v_{pinch}$ is an anomalous inward velocity typical in divertor studies, and $\eta_\perp$ is the anomalous cross-field viscosity.\cite{Stangeby_2000} Equation \ref{equation:MomBalance1} is complicated in its current state but can be simplified. \\

    The momentum source term $S_{mom}$ comes from particles that are ionized or from collisions with neutrals, i.e., $S_{mom} = S_{mom,n} + S_{mom,col}$. Neutral particles give momentum to the plasma if they already have momentum and take momentum away if they are generated with some momentum via recombination. This momentum source rate is expressed as
    \begin{equation}
    	S_{mom,n}(r,z) = m_i v_n S_p(r,z),
    \end{equation}
    where $S_p$ is the ionization source rate calculated in Equation \ref{equation:partbal}. Momentum transfer to or from neutral particles depends on the relative velocity of the plasma particles to the neutral particles. Since very little momentum is transferred between electrons and neutrals, we consider only ion-neutral collisions. The momentum gain is expressed as
    \begin{equation}
    	S_{mom,col}(r,z) = -\nu_{in}nm_i(v_{i,z}-v_{n,z})
    \end{equation}
    with collision frequency $\nu_{in}\approx500$ kHz for this plasma. We note that this term is positive, signifying a gain of plasma momentum when the neutral velocity is greater than the ion velocity, and negative, signifying a loss of plasma momentum when the opposite is true.\\
    
    We further simplify Equation \ref{equation:MomBalance1} by assuming because of the weak magnetization of ions and significant collisionality that there is no anisotropy in the parallel and perpendicular temperature in MAP, so $\pi_i$ can be ignored. The whole term $\left(D_\perp \frac{\partial n}{\partial r} + n v_{pinch}\right)$ is anomalous and is divided into two components for divertor research to identify the two separate contributions. These two terms together are identical to the total radial ion flux $nv_r$ measured directly on MAP. $\eta_\perp$ is taken to be of order $\eta_\perp\approx nm_iD_\perp$, following Stangeby, with $D_\perp\approx 3.5$ m$^{2}/$s. With these simplifications, we now have
    \begin{equation}\label{equation:MomBalance2}
    	\begin{split}
    		\frac{\partial}{\partial z}\left(p_i + p_e + nm_i v_z^2\right) & = -\nu_{in}nm_i(v_{i,z}-v_{n,z})\\
    		& + m_i v_n S_p(r,z)\\
    		& + \frac{1}{r}\frac{\partial}{\partial r}r\left(nm_i v_z v_r\right)\\
    		& + \frac{1}{r}\frac{\partial}{\partial r}r \left(nm_i D_\perp \frac{\partial v_z}{\partial r}\right).
    	\end{split}	
    \end{equation}
    We note that while Stangeby also describes contributions to momentum from the electric field, electron temperature gradient, and ion-electron collisions, these terms cancel when considering the total plasma momentum in an ambipolar plasma instead of considering the ion and electron momenta individually. For a full description of these equations, see Stangeby, Chapters 9.4 and 13.\cite{Stangeby_2000} \\

	\begin{figure}[htbp]
		\centering
		\includegraphics[width=\linewidth]{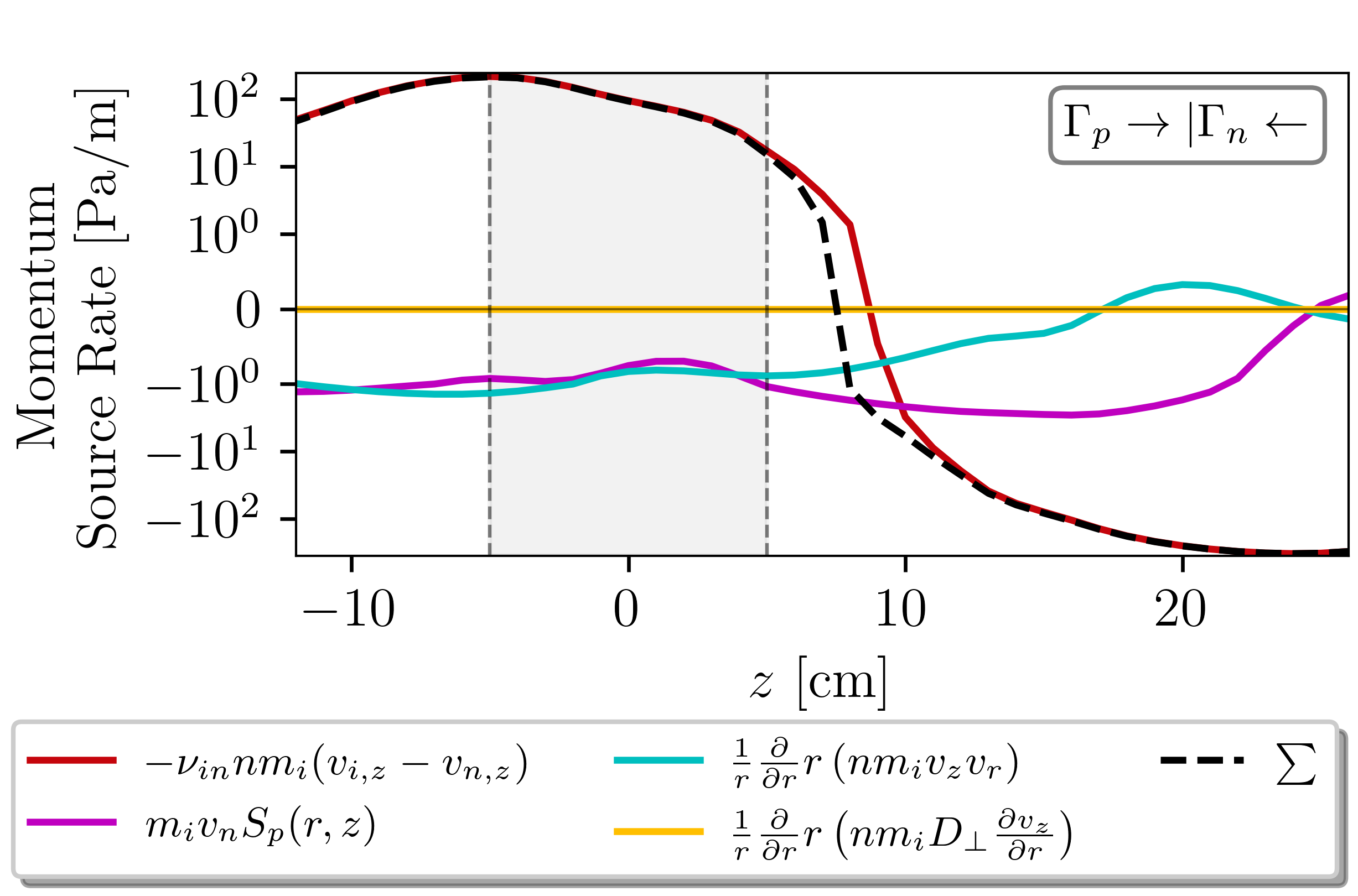}
		\caption{The momentum source rate for the antiparallel flow configuration is calculated from the right-hand side of Equation \ref{equation:MomBalance2}. The y-axis is logarithmic except from $-1$ to $1$ Pa/m, where it is linear. The antenna is located in the shaded region.}
		\label{fig:momA}
	\end{figure}

    Figure \ref{fig:momA} is an example of the individual components comprising the momentum balance. The y-axis is logarithmic except from $-1$ to $1$ Pa, where it is linear. The four terms of equation \ref{equation:MomBalance2} are plotted individually, and the sum is plotted in the dashed black line. The first term is due to the frictional momentum transfer from neutral particles to ions. This term is plotted in red and dominates throughout the plasma. The second term is due to neutral particles that are ionized with some initial momentum that is then transferred by the ionization into the plasma. This term is plotted in magenta. The third term is due to cross-field convection of momentum in the case of finite radial velocity gradients and is plotted in cyan. The fourth term is due to the cross-field diffusion of momentum and is plotted in yellow.\\

\begin{figure}[htbp]
		\centering
		\includegraphics[width=\linewidth]{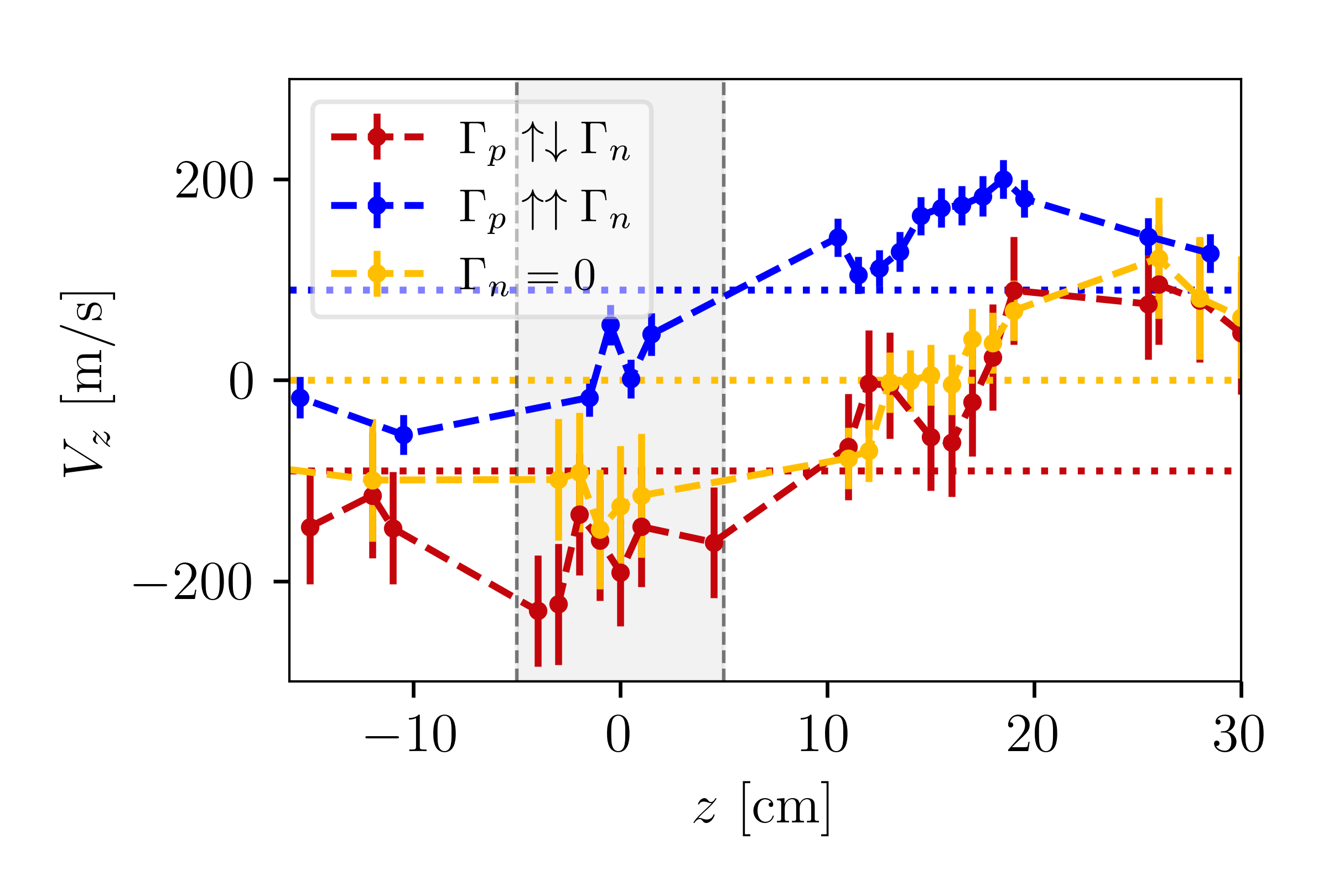}
		\caption{The axial ion velocity profiles for the antiparallel flow configuration (red), parallel flow configuration (blue), and no-flow configuration (yellow). The antenna is located in the shaded region. The calculated background neutral velocity is plotted as the horizontal dotted lines, colored to correspond with the respective configurations}
		\label{fig:VZAll}
	\end{figure}

    The relative axial velocity of ions and neutrals is a key parameter in the axial momentum balance. We plot the measured ion velocities in Figure \ref{fig:VZAll}. The neutral velocity is calculated from known device parameters. The true neutral velocity profile will be influenced by neutral depletion and interactions with the plasma, but we take the background neutral velocity as an estimate. The background neutral velocity is plotted in Figure \ref{fig:VZAll} in horizontal dotted lines, with colors corresponding to the respective colors of the plotted ion velocities measured with LIF. \\
    
    The most important feature of the relative ion-neutral velocity is that it crosses $0$ m/s. In the figure, this position is identified by the axial position at which the solid line crosses the correspondingly colored dotted line. To the right of the zero-crossing position, ions flow faster to the right than neutrals, and to the left of the zero-crossing position, ions flow faster to the left than neutrals. While this flow reversal immediately suggests an ionization source, it is also the strongest component of the frictional momentum loss term from Equation \ref{equation:MomBalance2}.\\
    
    The exact position of the crossing cannot be established for the configurations with neutral flow from the available data. The crossing occurs between $z=5$ and $z=10$ cm, but this also corresponds to a blind spot in the measurement because a magnetic field coil obstructs the collection optics line of sight. In the configuration without neutral flow, the crossing happens near $z=15$ cm, which is a measurable region, and is significantly farther downstream than the position of the zero-crossing in the other two configurations.\\
            
	\begin{figure}[htbp]
		\centering
		\includegraphics[width=\linewidth]{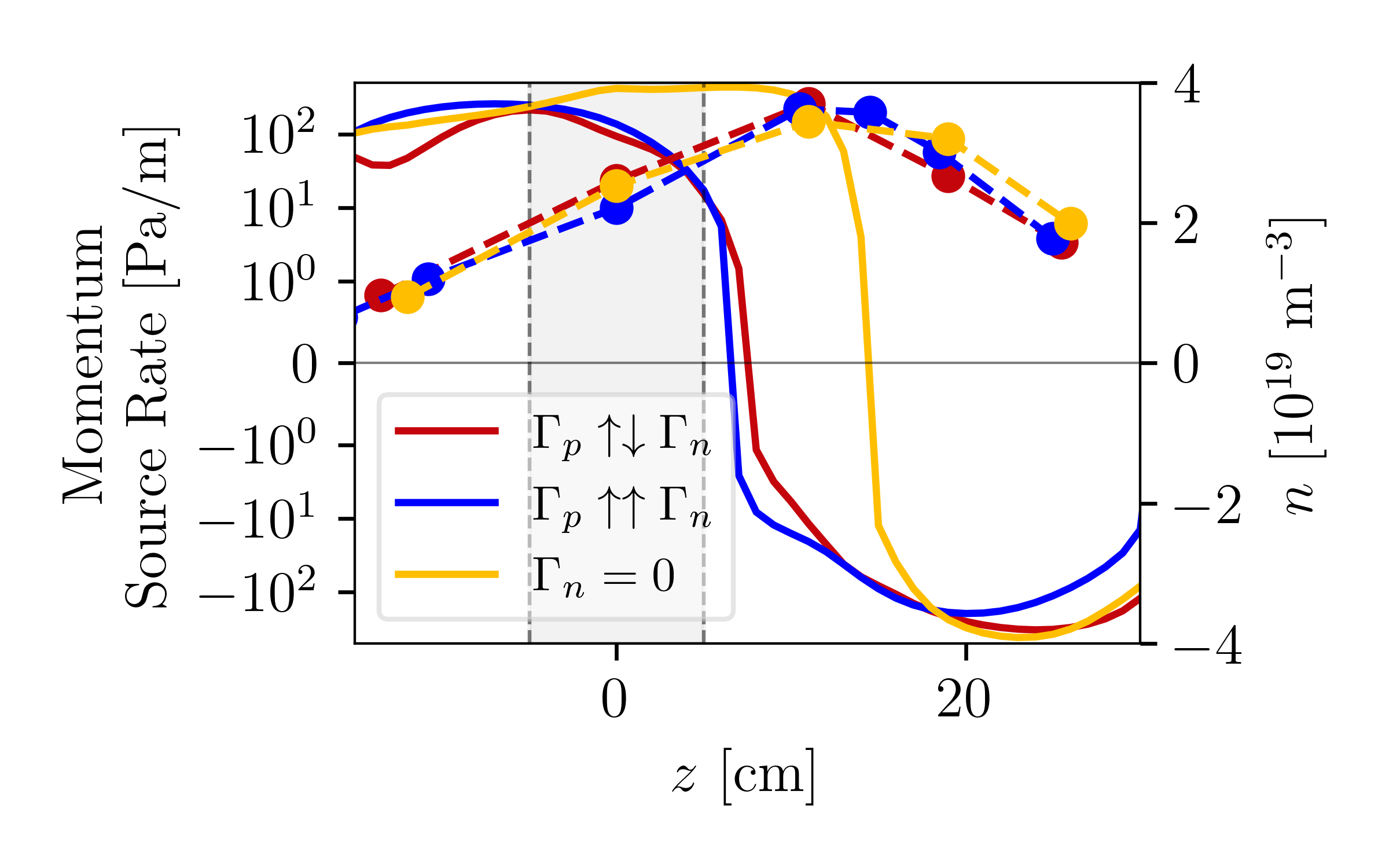}
		\caption{The momentum source rate for the antiparallel flow configuration (red), parallel flow configuration (blue), and no-flow configuration (yellow) as calculated from the right-hand side of Equation \ref{equation:MomBalance2}. The antenna is located in the shaded region. The left y-axis is logarithmic except from $-1$ to $1$ Pa/m, where it is linear. The density profiles from Figure \ref{fig:densityAll} are re-plotted on the right y-axis for comparison.}
		\label{fig:momAll}
	\end{figure}
    
    We plot the momentum source rate in solid lines for all three configurations in Figure \ref{fig:momAll}. We plot the density in dashed lines with colors corresponding to the colors used in previous figures to represent the three configurations. The momentum source rate profiles do not significantly vary between the configurations, aside from an axial shift. The shift corresponds to the shift in the zero-crossing in Figure \ref{fig:VZAll}.\\
    
    The plasmas in configurations with neutral flow gain rightward (positive) momentum at all negative $z$ but lose rightward momentum at positions starting before the density peak. This prevents ions from moving far to the right, as they do not maintain sufficient momentum to progress. The plasma in the configuration with no neutral flow is markedly different, as it maintains the positive momentum source rate until near the density peak. The plasma particles near the density peak in this configuration lose their outward momentum more slowly than in the other two configurations, thus allowing for a flattening of the axial density level. While the observed flattening of the profile was only slight, we have identified a mechanism in the momentum balance by which the flattening could be achieved.\\
    
    With more antennas and higher density, the ion-neutral friction will increase, so a stronger flow rate of argon will be required to fuel the plasma. For a very long plasma with many antennas, this flow would be impractical, so a pre-filled tube will be necessary, and radial recycling will still maintain the majority $(\approx90\%)$ of the refueling. The remainder $(\approx10\%)$ of the fueling may not be able to come from axial flow along the entire length of the tube, as the ion-neutral friction could prevent neutrals from passing more than a few antennas. We conclude that such a long plasma would benefit from fueling of additional argon along the length of the tube to maintain a homogeneous density profile. \\

	\begin{figure}[htbp]
		\centering
		\includegraphics[width=\linewidth]{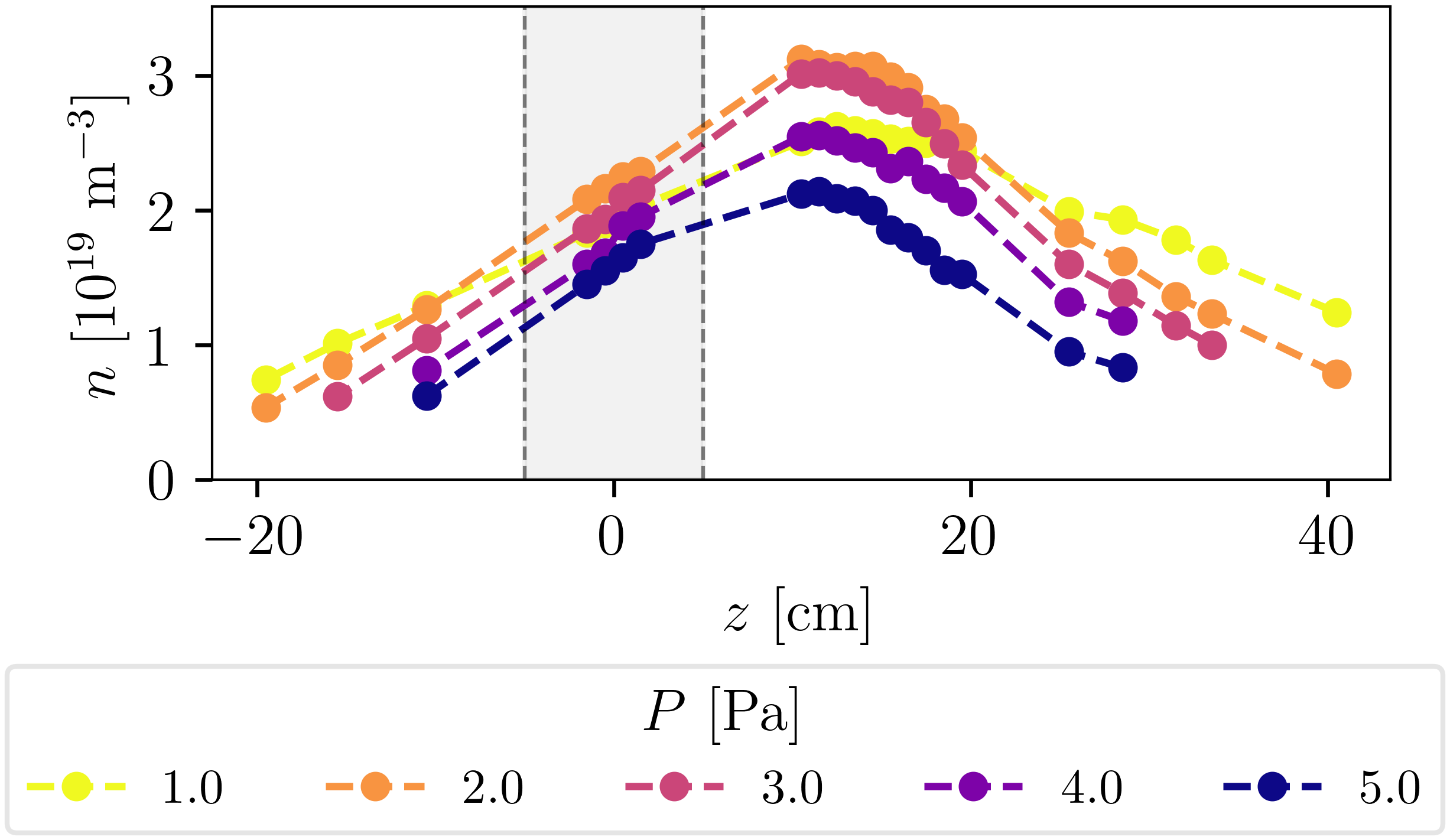}
		\caption{The axial density profile for the antiparallel flow configuration with varying background neutral pressure from $1$ Pa to $5$ Pa.}
		\label{fig:Pressure}
	\end{figure}

    Besides the flow of neutrals, the background neutral pressure provides a strong contribution to particle sourcing in plasmas. The neutral pressure in MAP was scanned from $1$ to $5$ Pa, and the axial density profile was measured at each pressure. The results are plotted in Figure \ref{fig:Pressure}. The density profile is notably widest for $P=1$ Pa, decreasing in width as the pressure is raised. From $P=1-3$ Pa, the plasma density increases or nearly maintain its peak value as the profile narrows. At pressure greater than $P=3$ Pa, the shape of the density profile remains nearly constant, and only the magnitude changes as the density decreases with increasing pressure.\\

    These results show the importance of balancing the available power density with the quantity of neutrals available for ionization. At low pressure the neutral density is so low that the helicon wave is able to support a significant plasma density far downstream from the antenna. At high pressure, collisions become more prevalent and restrict the motion of ions while also acting as a power dump that could result in particle heating instead of ionization. Between these two extremes there are sufficient particles present near the antenna to reach the maximum attainable density for helicon mode conditions, but not so many that too much power is lost to the neutral population. As the density is increased by the use of higher applied RF power, the pressure must be increased accordingly to provide a sufficient population for ionization and the efficient propagation of the helicon wave.  \\

    \section{Two Antenna Helicon Operation}\label{sec:twoant}
\begin{figure}[htbp]
		\centering
		\includegraphics[width=\linewidth]{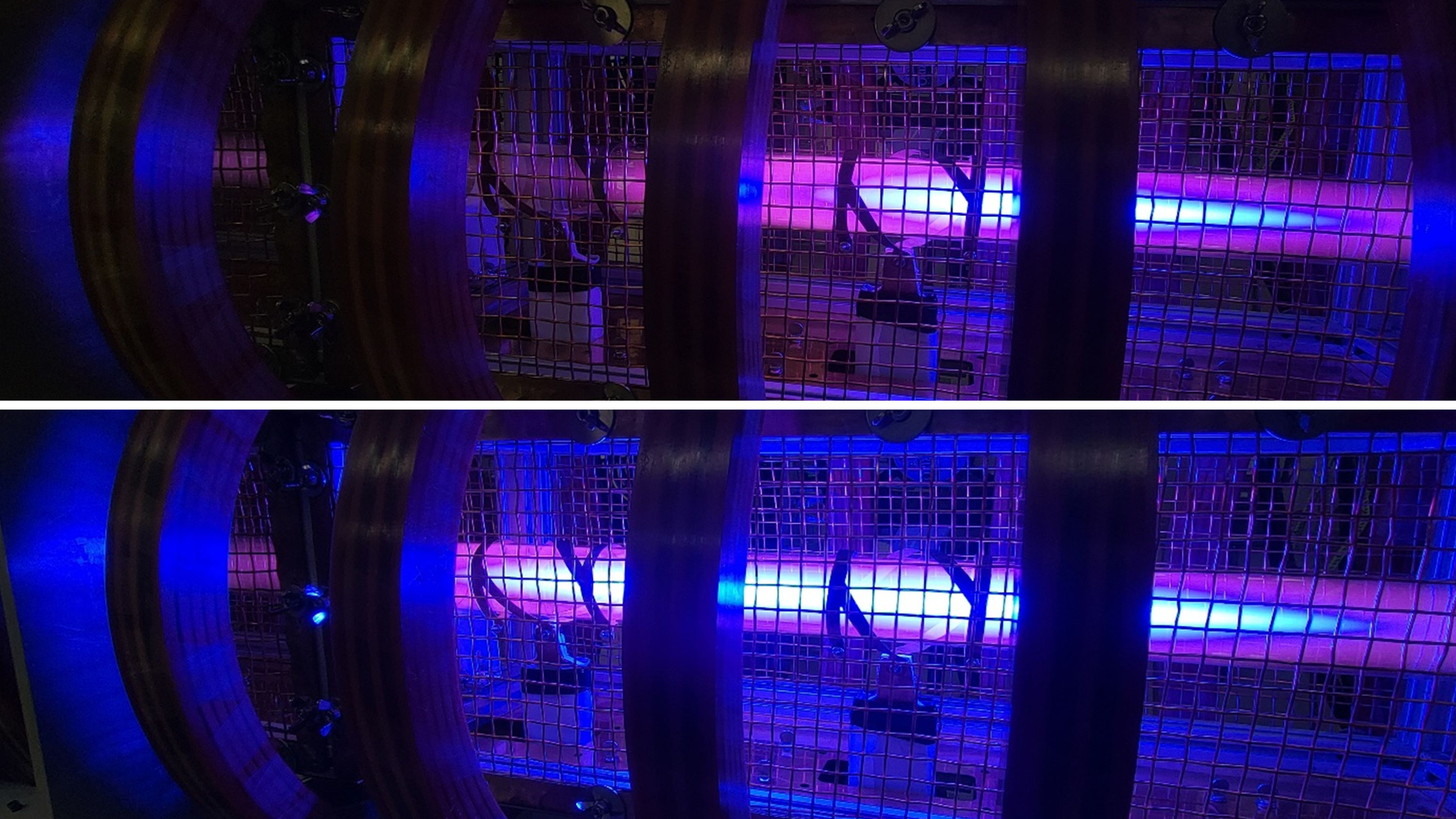}
		\caption{Helicon Plasmas in MAP are photographed in MAP. The bright argon plasma (blue and pink) is seen in the vacuum chamber, surrounded by two helical antennas. The Faraday cage and several of the $14$ electromagnets are seen in front of the chamber. MAP is depicted with plasma excited by a single antenna (top) and by two collinear antennas (bottom).}
		\label{fig:MAP}
	\end{figure}

\begin{figure}[htbp]
		\centering
		\includegraphics[width=\linewidth]{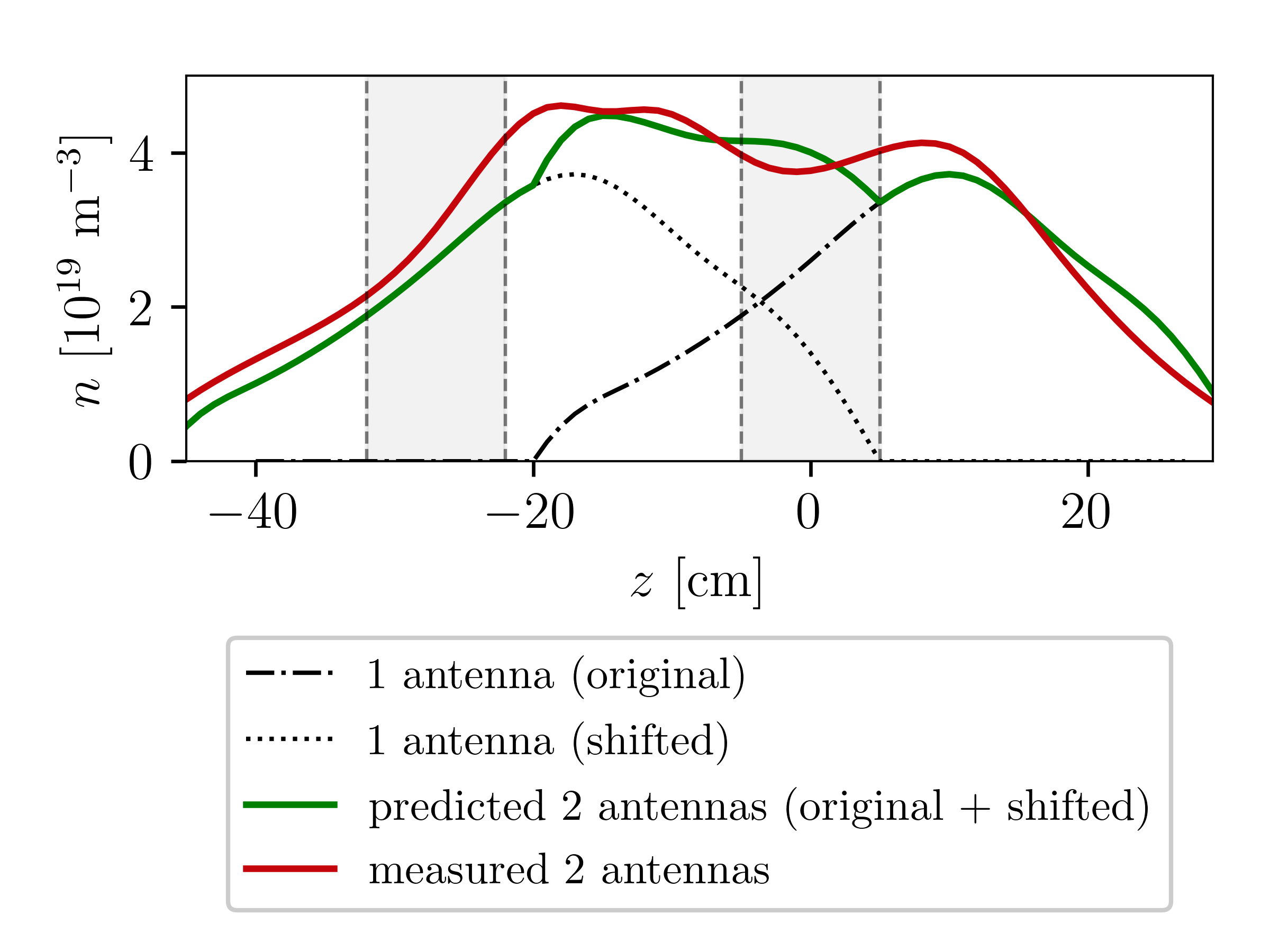}
		\caption{The axial density profile as measured in a single antenna plasma (black dot-dashed profile). The profile is shifted by $z=-27$ cm (black dotted profile). The two profiles are added (green profile) to predict the total density profile, which was then directly measured (red profile).}
		\label{fig:density_2ant}
	\end{figure}

    Having considered the density, velocity, ionization rate, and momentum source rate for three configurations of single-antenna plasmas, we now move on to examine how these results change in the case of plasmas excited by two identical antennas shifted from each other along the axis of the device by some distance $s$.  An example picture and density profile of a two antenna helicon are shown in Figure \ref{fig:MAP} and the red profile in Figure \ref{fig:density_2ant}, respectively. The top panel of Figure \ref{fig:MAP} shows a plasma excited by a single RF antenna at $1$ kW, while the bottom panel shows a plasma excited by two identical antennas at $1$ kW each. All other conditions are identical to those presented in Section \ref{sec:oneant}. \\

    We see from these two figures that the plasma maintains significant density for a much larger region along the axis. We will consider two mechanisms for the increased plasma length. First, the plasma could be longer because each antenna excites some plasma, and the two plasmas from the two antennas link to form a single plasma. In this case, we would expect a linear superposition of relevant parameter profiles from the two individual plasmas. Second, the plasma could be longer because the particles in the plasma formed near one antenna gain additional energy when they approach the other antenna, so they are able to progress farther along the axis without being lost. In this case, the relevant flows would gain new significance in determining the profile shapes.\\

    We consider various ways of identifying the cause of the longer plasma in a two-antenna configuration. First, we compare the two-antenna axial density profile to the sum of two single-antenna axial density profiles. The density profile measured in a plasma with a single antenna centered at $z=0$ cm is plotted in the dashed line in Figure \ref{fig:density_2ant}. The same profile is plotted again but shifted axially by $-27$ cm to correspond with the position of the second antenna in the two-antenna configuration. This profile is plotted in the dotted line. The sum of the dashed and dotted  lines is plotted in the solid green line, which represents the predicted two-antenna axial density profile for the case where the resultant density profile is a linear superposition of two single-antenna density profiles. The agreement of the red and green profiles in Figure \ref{fig:density_2ant} suggests that linear superposition of the density profiles does apply.\\

    \begin{figure*}[htbp]
		\centering
		\includegraphics[width=0.9\linewidth]{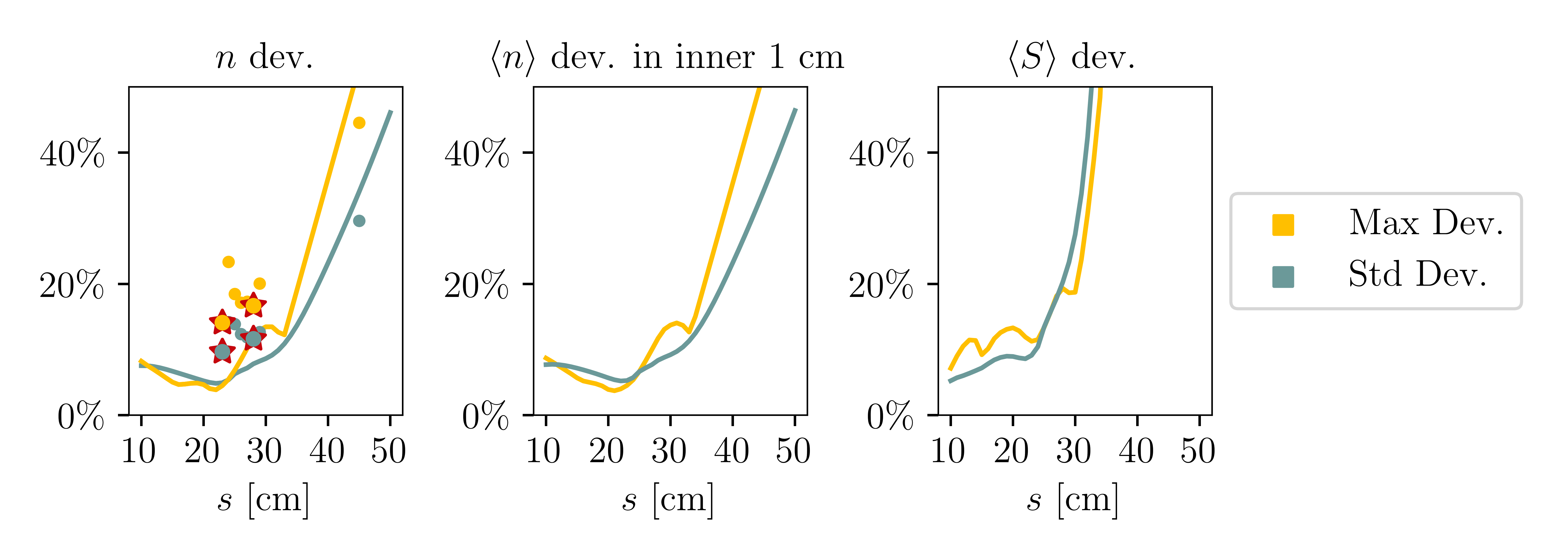}
		\caption{The calculated maximum (yellow) and standard (gray) deviations for the on-axis density (left), the average density in the inner $1$ cm of the plasma (middle), and the average ionization source rate in the whole plasma diameter (right) are plotted against antenna spacing $s$. Predicted values based on one-antenna plasma measurements are plotted as lines, and measurements from two-antenna plasmas are plotted as points. The two spacings with the lowest deviations are indicated by red stars.}
		\label{fig:spacing}
	\end{figure*}

    \begin{figure}[htbp]
		\centering
		\includegraphics[width=0.9\linewidth]{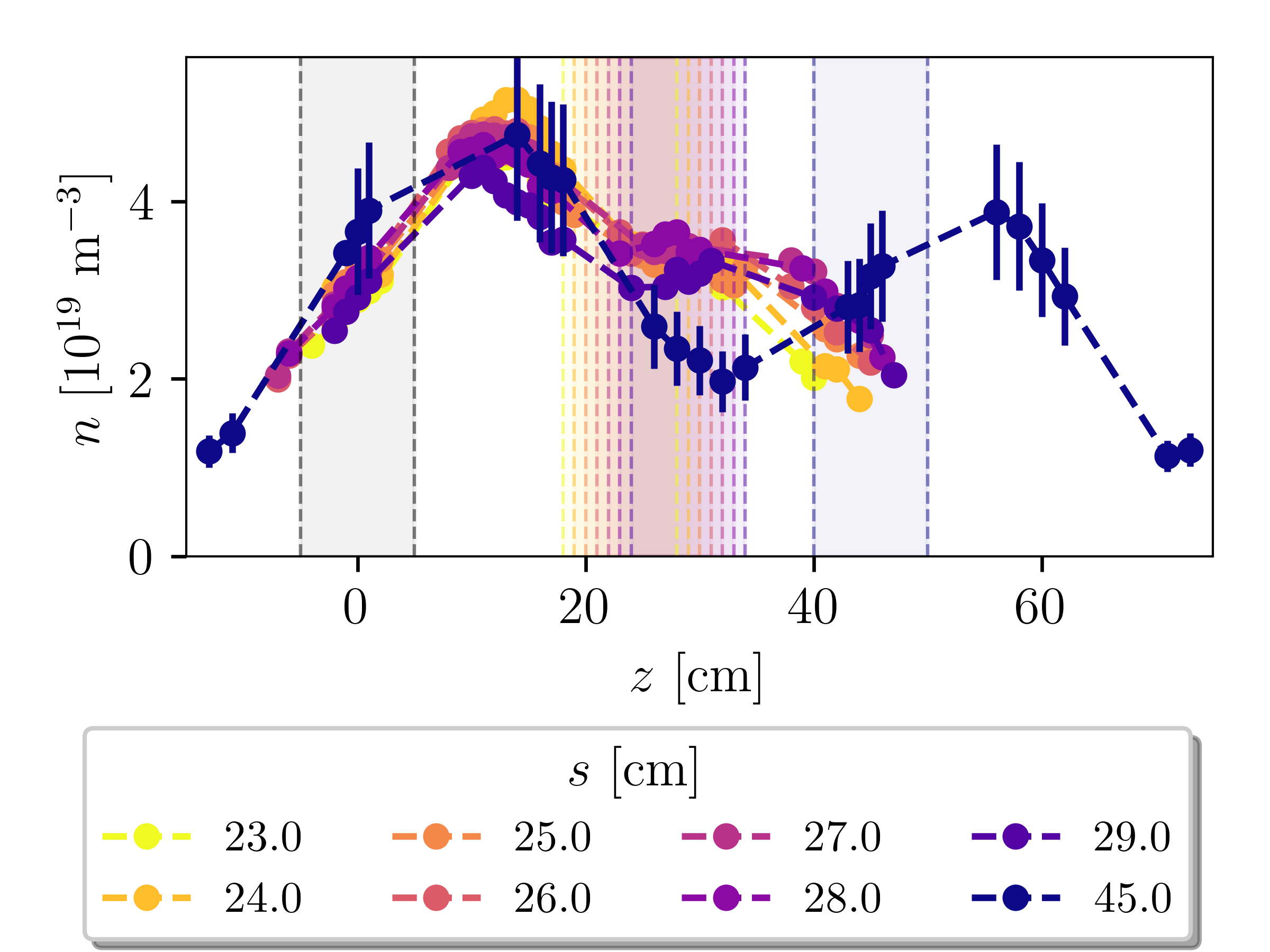}
		\caption{The axial density profile for the antiparallel flow configuration with varying antenna spacing. Typical uncertainties are shown on the $s=45.0$ cm trace. The left antenna is defined as $z=0$ cm. The right antenna (shifted at $s$) is indicated by the shaded region matching the color of the corresponding density profile.}
		\label{fig:spacing_data}
	\end{figure}
    
    Utilizing the near linear superposition of density profiles from multiple antennas, we calculated the expected change in the axial uniformity for various antenna spacings, defining the on-center distance between the two antennas as $s$. We calculate the maximum and standard deviations in the axial distributions from the edge of one antenna to the corresponding edge of the next antenna in a two antenna configuration, using the measured profiles from a single antenna profile and linearly superposing them upon each other. Results are shown in Figure \ref{fig:spacing}. In all panels, the lines represent predictions based on single antenna measurements, and the points represent measurements from a two-antenna plasma, as plotted in Figure \ref{fig:spacing_data} against the axial position. All measured profiles exhibit a double peak, with one peak to the right of each antenna. The deviations plotted in Figure \ref{fig:spacing} are calculated between the right sides of the two antennas. The deviations plotted in the left panel are calculated from the on-axis density, those in the middle panel are calculated from the average density in the inner $1$ cm of the plasma column, and those in the right panel are calculated from the average ionization source rate across the entire diameter. There is no significant relative change in the deviations between the three panels, except that the steep increase in deviation percentage is higher for the ionization source rate calculation than for either density calculation. \\
    
\begin{figure*}[htbp]
		\centering
		\includegraphics[width=0.9\linewidth]{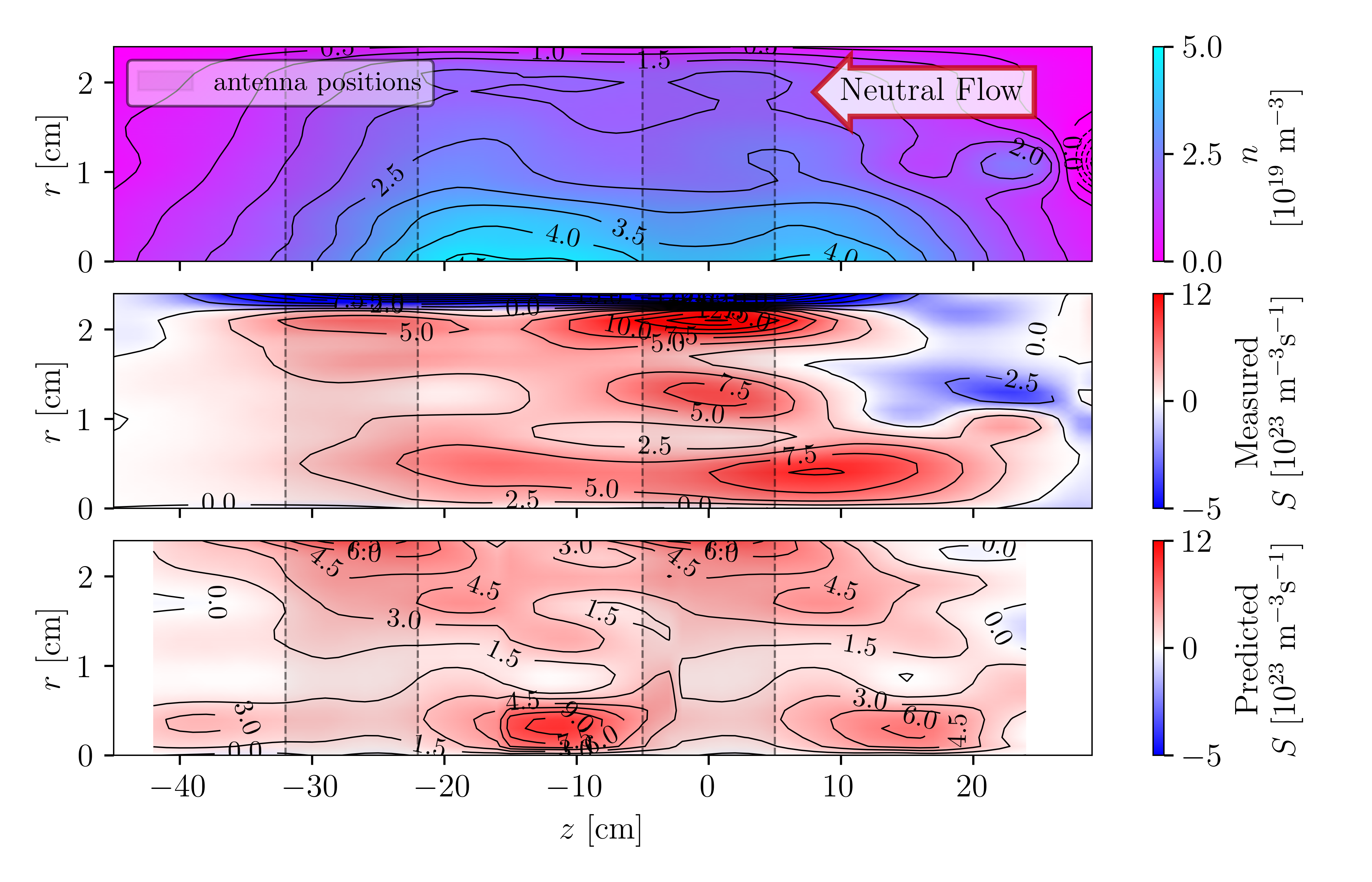}
		\caption{The measured $2$D density (top) and ionization source rate profiles (middle) and the predicted $2$D ionization source rate profile (bottom) are plotted. Predicted ionization source rate profile was taken from the linear superposition of two identical profiles, as in Figures \ref{fig:SR_2ant} and \ref{fig:spacing}.}
		\label{fig:SR_2ant}
	\end{figure*}

    We repeat the addition of profiles for $2$D ionization source rate profiles to explore to what extent the linear superposition applies. The results are shown in Figure \ref{fig:SR_2ant}. In the top panel, we plot the measured $2$D density profile, formatted as in Figure \ref{fig:densityAll} but with two antennas, one centered at $z=-27$ cm, and the other at $z=0$ cm. We observe a density peak between the two antennas of $n\approx4.6$ m$^{-3}$, and a separate peak to the right of both antennas of $n\approx4.1$ m$^{-3}$. The shape of the density profile is similar to the profiles plotted in Figure \ref{fig:densitycontourAll}\\
    
    For the same configuration, we plot the measured ionization source rate in the middle panel of Figure \ref{fig:SR_2ant}. To compare the measurements with the results predicted by linear superposition, we measured the ionization source rate in the same configuration but with a single antenna. We then shifted this ionization source rate profile as in Figure \ref{fig:density_2ant}, and the results are shown in the bottom panel of Figure \ref{fig:SR_2ant}. The antennas are indicated in all panels by the shaded boxes. The background neutral flow is from right to left, as indicated. \\
    
    The key features identified in Figure \ref{fig:SRAll} are also present in the middle and bottom panels of Figure \ref{fig:SR_2ant}, namely, the stronger ionization source rate directly under the antenna near the radial boundary and near the on-axis density peak. The most significant difference between the measured and predicted ionization rate profiles is the relative strength of the ionization source rate features near the right antenna and between the two antennas. The measured profiles show the highest ionization rate near $z=10$ cm, while the predicted profiles show the highest ionization rate near $z=-10$ cm. The predicted profile peak is highest between the antennas because there is some component from each antenna contributing to the net ionization rate in this inter-antenna region, while the peak near $z=10$ cm in this case does not receive any contribution from the antenna centered at $z=-27$ cm.\\

    The cause of the rightward shift of the highest ionization rate in the measured two-antenna profiles is less apparent. Any cause of the altered ionization rate profile must be small enough to not significantly alter the density profiles from a linear superposition. The density profiles are determined based on fueling, but also based on transport after the ions are sourced. The transport must adjust adequately to maintain the linear superposition of two single-antenna density profiles. This suggests that the geometry of the device could have a more significant impact on the density profiles than the ion sourcing has. The ion sourcing could, however, be influenced significantly by the neutral argon as it flows through the chamber. While the majority of the plasma fueling is from radial recycling, there is also a supply of neutrals that is refreshed on the right side of the plasma, as argon flows into the chamber. If the population of neutral argon available for ionization is depleted near the ionization source rate peak on the right side of the plasma, there may be insufficient neutrals reaching the ionization source rate peak between the antennas, so the helicon is unable to sustain the peak ionization rate in that region. In that case, we conclude that the ions generated near the right antenna must progress from the right antenna toward the left in a sufficient quantity to maintain the peak density in the region between the two antennas. \\

\begin{figure}[htbp]
		\centering
		\includegraphics[width=\linewidth]{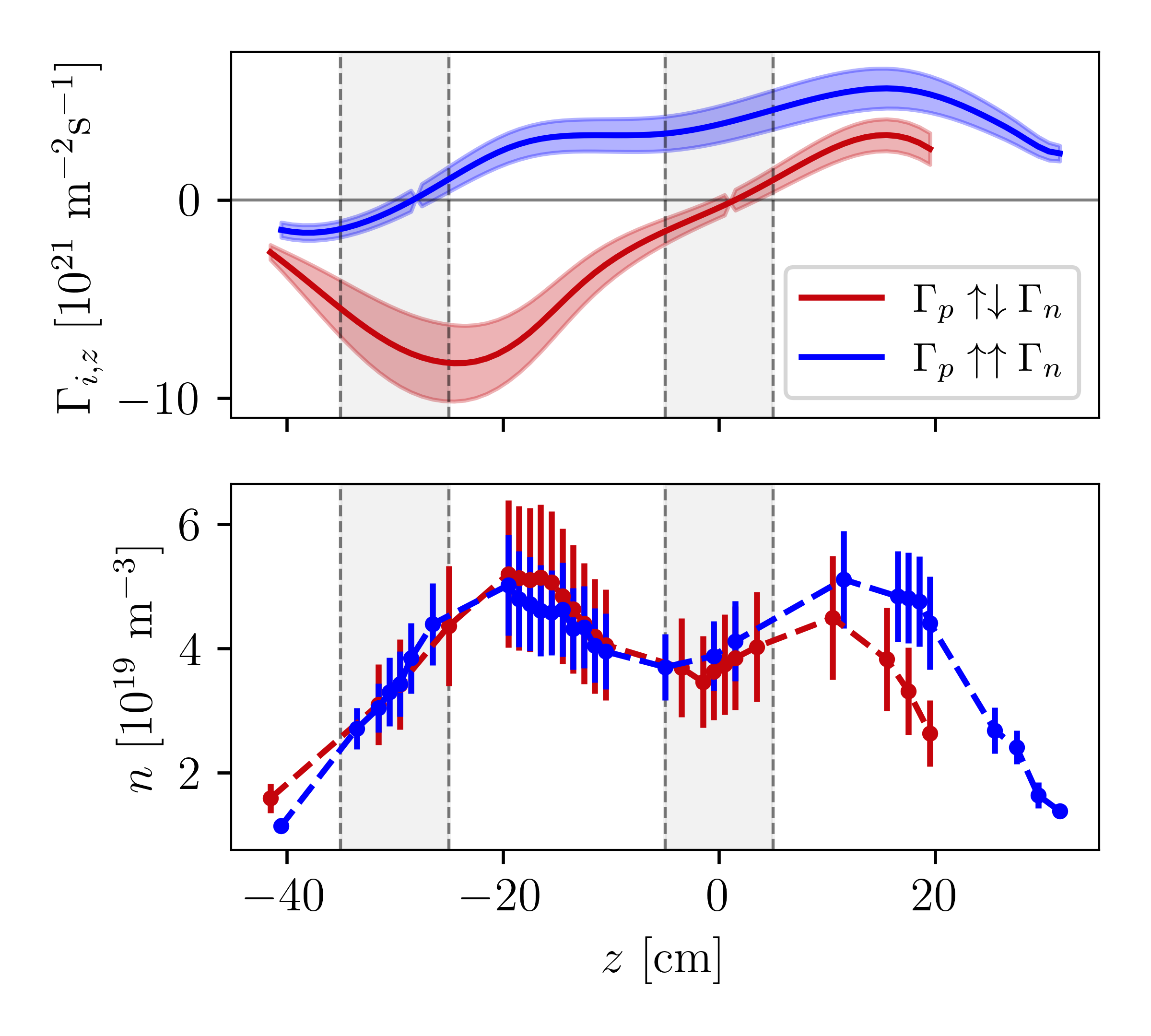}
		\caption{The axial ion flux (top) and density (bottom) profiles for parallel (red) and antiparallel (blue) plasma and background neutral flow directions in a two-antenna plasma with $s=30$ cm.}
		\label{fig:FluxComparison}
	\end{figure}

    To explicitly quantify how the change in neutral flow influences the plasma, we measured the axial ion flux and the axial density profiles with parallel and antiparallel background neutral and plasma flows.  The device setups used to acquire this data are identical to the antiparallel and parallel flow configurations, but with the installation of a second antenna $30$ cm from the original antenna. The results are plotted in Figure \ref{fig:FluxComparison}. The ion flux profiles in the top panel agree only to first order between the antennas, each increasing in the positive direction (right, direction of plasma flow). There is also a step-like characteristic to each profile, as there is a large gradient near the antennas with a flatter region between them. This larger gradient near the antenna corresponds with the flow reversal observed in the one-antenna plasmas.\\

    The density profiles plotted in the bottom panel of Figure \ref{fig:FluxComparison} also exhibit first order agreement. The most significant difference between the two occurs in the density peak near $z=15$ cm. In this region, the peak is notably higher in the configuration with neutrals flowing from left to right (parallel flows) than in the converse configuration. In the antiparallel configuration (red), ions generated near the right antenna are generally lost to the right, as the flow reversal point is under the antenna, while the density peak is to the right of the antenna. In the parallel configuration (blue), ions generated near the left antenna flow to the high density region of the right antenna. This could be an effect of the neutral flow from left to right, if the neutral argon carries the ions along with it through collisions. The result of these flows is that the density peak between the antennas is primarily caused by the ionization of the left antenna in both configurations. The peak to the right of the right antenna is primarily caused by the ionization of the right antenna in the antiparallel configuration, but both antennas contribute substantially in the parallel configuration.\\
    
    To expand this investigation with two antennas, we measured the axial density profile in four plasma configurations. The first three configurations are those considered in Section \ref{sec:oneant}. The fourth configuration is a variation of the no-flow configuration. This new configuration is still excited in a background neutral population with no flow, but the plasma is launched toward the open pumping region on the left end of the device, instead of towards the closed end cap on the right end of the device. \\
    
    \begin{figure*}[htbp]
		\centering
		\includegraphics[width=0.9\linewidth]{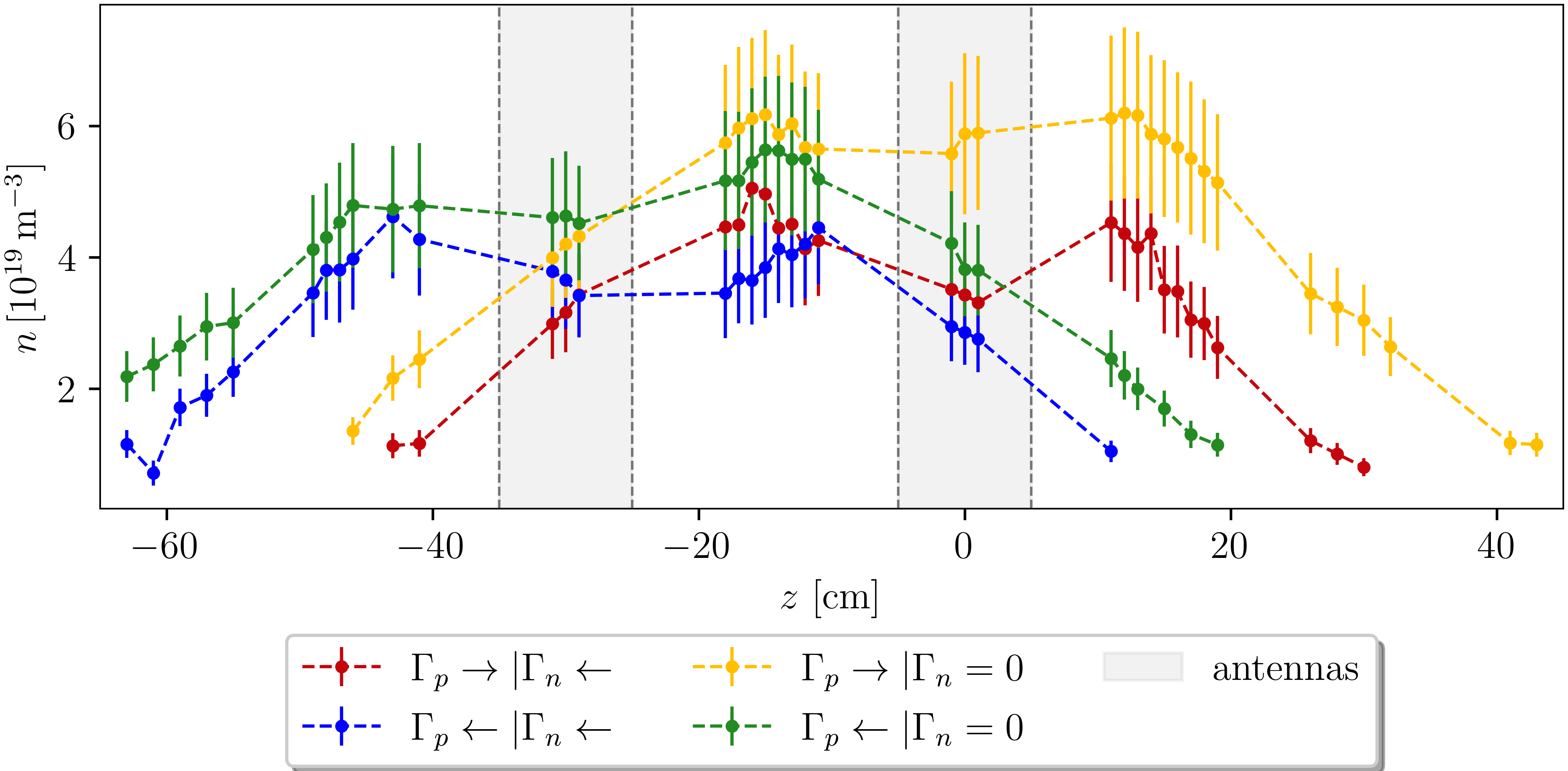}
		\caption{The measured density profiles for plasma launched from left to right (red and yellow) and right to left (blue and green), for neutral flow from right to left (red and blue), and no neutral flow (yellow and green). The red, blue, and yellow data correspond with the colors used in previous plots.}
		\label{fig:flows}
	\end{figure*}

    We plot the four density profiles for two antenna plasmas in Figure \ref{fig:flows}. The antennas are spaced $30$ cm along the axis. The red, blue, and yellow curves are defined as above, except that we now use the physical position along the MAP device as the x-axis, and do not flip the profiles to all overlap on the same side of the plot. The configurations with neutral flow (red and blue) are nearly mirror images of each other with one important exception, which we discuss below. The configurations with no neutral flow exhibit considerably higher densities than the configurations with neutral flow. The configuration with no neutral flow and with plasma flow to the left (green) is surprisingly different from the corresponding configuration with plasma flow to the right (yellow), as the latter has a higher density and equally high peaks, while the former has a lower density and one peak that is considerably higher than the other. The lower peak is likely due to the proximity to the pumping region, at which neutrals are replenished, although not as quickly as in the configurations with neutral flow through the main chamber.\\

    In comparing the configurations with background neutral flow, we identify an effect that the neutrals have on the density profile in a two-antenna helicon. Both configurations exhibit higher density peaks on the downstream neutral flow side than in the upstream side. This is most apparent in the case of antiparallel plasma and neutral flows (red), in which the peak between the antennas is $\approx 10\%$ higher than the peak to the right of the right antenna. This difference is expected to be a result of the neutral flow causing ions to flow from right to left, thus leading to a higher density on the left side.\\

    \begin{table}[htbp]
        \centering
        \begin{tabular}{|c|c| c|}
            \hline
             & Standard Deviation & Maximum Deviation \\
            \hline
            $\Gamma_p \rightarrow | \Gamma_n \leftarrow$ & $11\%$ & $21\%$ \\
            $\Gamma_p \leftarrow | \Gamma_n \leftarrow$ & $11\%$ & $17\%$ \\
            $\Gamma_p \rightarrow | \Gamma_n=0$ & $5\%$ & $8\%$ \\
            $\Gamma_p \leftarrow | \Gamma_n=0$ & $5\%$ & $12\%$ \\
            \hline
        \end{tabular}
        \caption{Standard and maximum density deviations from data in Figure \ref{fig:flows}. Deviations are calculated between the edges that are downstream in the plasma flow direction for each antenna.}
        \label{table:2}
    \end{table}

    We calculate the standard and maximum deviations of the densities plotted in Figure \ref{fig:flows} between the downstream (in the plasma flow) edges of the two antennas. The results are shown in Table \ref{table:2}. We observe that the direction of a background neutral flow does not significantly alter the homogeneity. However, there is a significant improvement in homogeneity when the neutrals do not have a predefined background flow through the chamber. The homogeneity improves by approximately a factor of $2$ according to both metrics.\\

    We paint a possible picture of what the neutrals see in either configuration when they are injected into the plasma. For antiparallel flow, the neutrals injected into the chamber are met with ions from the plasma, streaming towards them. The ion flux pushes back on the neutrals, but as the neutrals already have their own momentum and more neutrals are constantly pumped in from behind, the neutrals must keep moving. The high ion pressure and short ionization mean free path on the axis of the device likely results in many neutrals moving along near the outer radius of the device, in accordance with the on-axis depletion of neutral particles commonly observed in helicons.\cite{Gilland_1998,Fruchtman_2005,Magee_2013} As the rest of the neutrals continue along the device, many will diffuse radially towards the axis of the device and be ionized there, eventually moving back towards the gas injection position or recycling again radially. This radial movement establishes about $90\%$ of the particle balance as discussed before and the axial flow therefore deals with the refueling of any lost particles along the axis. When the remaining neutrals reach the area under the antenna, they are confronted with a second region of large power deposition and, hence, another position with a significant ionization source rate. This is due to the presence of the TG mode directly under the antenna. Many of the neutrals in this region are ionized. The neutrals that remain will continue moving along the device at the outer radius and are met by another region of power deposition from the left antenna. This will cause more of the neutral particles to be lost as ions. At this point, there are far fewer neutral particles left, so when the remaining neutrals reach the left antenna and the region under it where the TG mode propagates, there are not many particles left to be ionized. What particles make it through are pumped out on the left end of the device. These, however, are mostly particles that are not ionized at the last antenna, because the ionization scale length in these plasmas is much shorter than the device length. The scale length will shrink further with the AWAKE relevant densities. \\
    
    Overall, the neutral movement is dominated by local radial movement into an ionization region, but there is some fraction of neutrals that is able to progress in the axial direction for a significant distance before moving inwards. There are also ions that move outwards radially and recombine to form neutrals again, and these neutrals can also progress in the axial direction. The local radial recycling supports the global axial movement of neutrals, but without maintaining the original neutral flux. The axial flux is lost from the plasma at the axial limit, where ions leave the plasma and are not recycled, but are replenished by neutrals injected from the opposite axial limit.\\
    
    The picture is almost identical for parallel flow, with only a few key differences. Neutrals from the right are met with ions streaming towards them, as before, but these ions have less relative momentum in this case because they were formed from a gas that already had a leftward momentum. Because of this lower momentum difference to the right, not as many neutrals are pushed back towards the injection point, so there is not as severe of an increase in pressure on the gas inlet side. Most of the neutrals continue moving left, as above, but they continue to add to the general leftward motion instead of fighting it. This allows for easier fueling of the plasma near the left antenna since there is no significant pressure drop on the left side. The result is a more uniform profile, with the density peak caused by the second antenna receiving nearly as much fueling as the density peak caused by the right antenna and also receiving any ions that make it far enough left without being lost to recombination.\\

    \section{Conclusions}\label{sec:conclusion}

    We have manipulated the background flow of argon through a helicon plasma to study the impacts of neutral fueling on the particle balance. Neutral fueling contributes to the resultant density and temperature profiles, and so it is important to understand when particular parameters are required for an application. We have utilized both one-antenna and two-antenna helicons in these measurements to expand the scope of the study to include results regarding how the plasma generated by each antenna influences the fueling of the plasma.\\

    The results presented here show that the background neutral flow does not significantly influence the density profile in a one-antenna plasma, as there is only a slight flattening of the axial density profile in configurations with no background neutral flow. In a two-antenna plasma, however, configurations with no background neutral flow exhibit a much more pronounced flattening as well as an increase in the density. Through the derived axial momentum balance, we concluded that the frictional losses from ion-neutral collisions are the dominant momentum loss channel. The relative ion-neutral velocity profile is significantly altered in configurations with no background neutral flow, likely leading to the change in the axial density profile.\\
    
    For use in particle accelerators, we suggest the use of a gas injection configuration with no neutral flow through the chamber, as this configuration leads to a factor of two improvement in the axial density uniformity and a $50\%$ increase in absolute density. If additional particles are needed to sustain high power operation, a higher pressure should be implemented before changing the injection scheme. To counter axial losses, it may be necessary to inject particles through valves distributed along the axis of the accelerator plasma, especially between plasma pulses if axial losses have caused a significant pressure buildup at one end. The axial losses could induce an axial density gradient which must be balanced by some moderate local gas injection at each or every other antenna to maintain the density homogeneity. It was shown that these axial losses address about $10\%$ of the total number of particles that are ionized, but the sensitivity of the density profiles to this external particle source have highlighted the need for such an active fueling scheme to minimize global flow, rather than injecting neutrals at one end and allowing them to flow directly through the device. The active fueling would also provide an additional free parameter for control of the axial density profile. Active fueling can be used in conjunction with tuning of the RF power, antenna spacing, and local magnetic field strength to optimize the axial density profile so that it can reach the AWAKE requirements.\\

    \section*{Acknowledgments}
    The research presented here was funded by the National Science Foundation under Grant Nos. PHY-1903316 and PHY-2308846 and the College of Engineering at the University of Wisconsin--Madison through the Thomas and Suzanne Werner Professorship of the Department of Nuclear Engineering and Engineering Physics.

    \nocite{*}
    \bibliography{references}
	
\end{document}